%% file: RambaldiFilimonovLillo.tex
\documentclass[1p,preprint, authoryear]{elsarticle}
\usepackage[utf8]{inputenc}
\usepackage[english]{babel}
\usepackage{amsmath}
\usepackage{amsfonts}
\usepackage{amssymb}
\usepackage{graphicx}
\usepackage[round]{natbib}
\usepackage{booktabs}  % better tables
\usepackage{makecell}
\usepackage{eurosym}

\usepackage{bm}% permette di usare grassetto nelle formule matematiche
\usepackage{bbm}% è simile ma permette di fare 1 in blackboard bold

\usepackage{amsthm} %theorems

\usepackage{mathtools} %allows cases definitions for equations
\usepackage{multirow}

\usepackage[footnotesize,bf]{caption} %pacchetto per controllare lo stile delle capition
\usepackage[section,above,below]{placeins} %allows to prevent floats being positioned after a \FloatBarrier command
\usepackage{hyperref}

\newcommand{\diff}{\mathrm{d}}
\newcommand{\E}[1]{\mathbb{E}\left[ {#1}\right]} %expected value

\newcommand{\fil}{{\mathcal{F}}} %filtration symbol
\newcommand{\like}{\mathcal{L}} %likelihood symbol
\newcommand{\AIC}{\mbox{AIC}}
\newcommand{\BIC}{\mbox{BIC}}
\newcommand{\almgtime}{\kappa}
\newcommand{\Nshock}{S}
\newcommand{\sigmaloc}{\sigma_{\mbox{\tiny loc}}}
\DeclareMathOperator*{\argmax}{arg\,max}
\DeclareMathOperator*{\argmin}{arg\,min}

\graphicspath{{./Images/}}

\usepackage{color}

\newtheoremstyle{no_par}% name of the style to be used
  {\topsep}% measure of space to leave above the theorem. E.g.: 3pt topsep is the same as in plain style
  {\topsep}% measure of space to leave below the theorem. E.g.: 3pt
  {\itshape} % name of font to use in the body of the theorem
  {} % measure of space to indent
  {} % name of head font
  {} % punctuation between head and body
  {5pt plus 1pt minus 1pt} % space after theorem head; " " = normal interword space. \newline for a line break !DOESN'T work with enumerate or itemize! to get a newline after the theorem head when using "eumerate", use \mbox{} after \begin{theorem}. 
  {\thmname{\textbf{#1}}\thmnumber{ \textbf{#2.}}\thmnote{ \textit{#3}}} % Manually specify head

\theoremstyle{no_par}

\theoremstyle{plain} %standard style

\theoremstyle{plain} %standard style

\journal{arXiv}

\addto\captionsenglish{}

\begin{document}

\begin{frontmatter}

\title{Detection of intensity bursts using Hawkes processes: \\an application to high frequency financial data}

\author[sns]{Marcello Rambaldi \corref{cor1}}
\ead{marcello.rambaldi@sns.it}

\author[eth,pr]{Vladimir Filimonov}
\ead{vfilimonov@ethz.ch}

\author[sns]{Fabrizio Lillo}
\ead{fabrizio.lillo@sns.it}

\address[sns]{Scuola Normale Superiore, Piazza dei Cavalieri 7, Pisa 56126, Italy}
\address[eth]{Department of Management, Technology and Economics, ETH Zurich, Switzerland}
\address[pr]{Department of Economics, Perm State University, Perm, Russia}
\cortext[cor1]{Corresponding author}

\begin{abstract}
Given a stationary point process, an intensity burst is defined as a short time period during which the number of counts is larger than the typical count rate. It might signal a local non-stationarity or the presence of an external perturbation to the system. In this paper we propose a novel procedure for the detection of intensity bursts within the Hawkes process framework. By using a model selection scheme we show that our procedure can be used to detect intensity bursts when both their occurrence time and their total number is unknown. Moreover, the initial time of the burst can be determined with a precision given by the typical inter-event time. We apply our methodology to the mid-price change in FX markets showing that these bursts are frequent and that only a relatively small fraction is associated to news arrival. We show lead-lag relations in intensity burst occurrence across different FX rates and we discuss their relation with price jumps.
\end{abstract}


\end{frontmatter}
\setlength{\parindent}{10pt}

%================================================================================ 
%================================================================================ 
\section{Introduction}

The detection of anomalous dynamics and regime changes is an important problem which has a wide range of applications in many systems from risk management to preventing of fraud. Financial markets are emblematic in this respect. Market participants, and particularly those who act as intermediaries, need to be resilient to sudden changes in market conditions that can arise endogenously or when a new piece of information becomes available. However, the identification of anomalous dynamics might be challenging, especially when the ``normal" dynamics is complex, possibly with non-linear and/or long range correlations. In finance, for example, a large amount of research has been devoted to the detection and characterization of ``anomalous'' price changes. Given an underlying price dynamics described by a continuous semimartingale, the problem is to test for the presence of price discontinuities (jumps - see for example \cite{ andersen2007no, Lee:2008ip, bollerslev2009discrete, ait2009testing} to cite only a few). 

In many real systems the observed time series is described by a point process. Examples include the arrival of phone calls, email, tweets, customers, queries, etc., and of course also financial time series, since at the smallest time scale the change in price is generically described by a (marked) point process. To the best of our knowledge the identification of anomalous changes in the intensity of a point process has remained relatively unexplored, especially when the ``normal" dynamics presents correlations and bursts of activity. In this paper we propose a method for tackling this problem. Specifically, we aim at identifying abrupt increases of the intensity of a otherwise stationary, yet correlated and bursty, point process, and where these increases relaxes back to the normal state after a certain period of time. We will call such events \emph{intensity bursts (IBs)}.
Detecting IBs might be important to identify anomalous activities in the system, detect the arrival of external perturbations, analyze the contagion effects and lead-lag relations in different co-evolving systems.

We propose a parametric model-based approach for the detection of intensity bursts, which relies on the Hawkes process~\citep{hawkes1971}. Being closely related to branching processes~\citep{Harris_Branching2002}, the Hawkes model combines in a very natural way external (exogenous) influence on the system with internal (endogenous) self-excited dynamics. Here the probability of arrival of a new event is given by the combination of a baseline probability and a contribution from all the former generated events. In natural and socio-economic systems,  Hawkes processes have become a very popular tool due to their flexibility and, at the same time, the simplicity of the calibration procedure. They have been applied in a variety of research domains such as seismology \citep{ogata1988statistical}, genomics \citep{reynaud2010adaptive}, neurophysiology (\cite{bremaud_stability, Chornoboy1988}) as well as in works on the spread of crime and violence \citep{lewis2012self,mohler2012self} and on social network dynamics \citep{SornetteCrane2008, NIPS2012_4834, zhou2013learning}.
Within the domain of financial applications, the Hawkes process has become a widespread model for the dynamics of high-frequency price changes and order book evolution~\citep{bacry2011, bormetti_cojumps, bouchaud_hawkes, filimonov2012quantifying, Bowsher2007, embrechts2011multivariate, Blanc:2015ua} and also was extended to the modeling of the price shocks on a daily scale (see for instance \citet{bauwens2009_review} and \citet{bacry2015hawkes} for a review).

We choose to model the ``normal" dynamics of the point process with a Hawkes process, so that correlations and bursts are present. On top of this we assume that in the process there are few IBs that locally perturb the counting dynamics and, via the excitation mechanism of the Hawkes process, affect also the ``normal" dynamics. In this setting, the parameters of the model and the location of the IBs are not known and must be inferred from the data. 

In \citet{Rambaldi2015} two of us have considered the related but much simpler problem of inferring the parameters of the model when the location of the IBs is known. The considered example was the arrival of scheduled macro-economic announcements in financial markets. These events have a major and dramatic impact on both price and trading activity and cannot be described within the classical Hawkes model, but the their timing is known in advance.
In the general case, the occurrence and timing of IBs are not known. The major limitation of~\citet{Rambaldi2015} is precisely the assumption that the occurrence time of the IBs is known, which prevents the study of unexpected (surprise) IBs.
Interestingly, even when the expected time is known in advance, a delay or anticipation of the actual arrival of the external event can bias the procedure. 

In the present work, instead, we consider the general case when both IBs arrival times and the total number of IBs are unknown. We propose an efficient and robust procedure of estimation from the empirical data, and we also suggest a hypothesis test that distinguishes the genuine intensity burst from a statistical fluctuation that could be equally well explained within a standard self-excited dynamics. 
We validate our procedure and show that it is capable of reliably identifying sudden increases in the event rate with a relatively low false positive rate. 
Furthermore, being a model-based, our approach allows to estimate parameters of the intensity burst together with the properties of the underlying self-excited process, and use it for classification of the anomalous activity.

We apply our approach to the analysis of high-frequency financial data from a major FX market. We identify a large number of IBs and we compare them with price jumps and macro-economic announcements. Interestingly, we find a large number of IBs which are not explained (in terms of time proximity) by neither of these two possible causes. This is noteworthy, since it signals the presence of different market anomalies and we propose possible explanations for them.

We foresee several major directions along which our method could be useful. First, the detection of an anomalous an unexpected market activity is essential for monitoring liquidity and for intra-day risk management applications. Moreover, the analysis of the historic data is essential for forensic investigations and detection of fraud and market manipulations. Besides, our method is relevant even when the analysis of the intensity burst itself is not the primary interest, and the focus is on describing the underlying process. Indeed, the procedure we propose mitigates Hawkes process tendency to overestimate the degree of self-excitation when burst-like non-stationarities are present. Finally, we stress that the potential range of applications spans well beyond the financial markets. For example our burst detection method could be of great importance in the analysis of social media dynamics such as YouTube views~\citep{SornetteCrane2008} or twitter posts~\citep{MacKinlay:2015wg}, or even in the analysis of data of social conflicts and unrest~\citep{Donnay:2014ie}.

The remaining of the paper is structured as follows. We start with the description of the Hawkes model that contains exogenous intensity bursts in Section~\ref{sec:model}. Section~\ref{sec:identification} presents the methodology for the statistical identification of intensity bursts. We test and validate our procedure on synthetic data in Section~\ref{sec:model_validation}. Section~\ref{sec:fx} presents an application of our methodology on real financial data. We conclude in Section~\ref{sec:conc}.
 
%================================================================================ 
%================================================================================ 
\section{Hawkes process with an exogenous intensity burst}\label{sec:model}

A univariate point process is a sequence of events that occurred at random times $t_i$ and that is described by the corresponding counting process $N_t=\sum\mathbbm1_{t_i<t}$, where $\mathbbm1_A$ is the indicator function of the set $A$. In the homogeneous Poisson process the intensity of the events is constant  $\lambda(t) = \lim_{\Delta\to 0} \Delta^{-1} \E{ N_{t+\Delta} - N_t }=\mu$.

\citet{hawkes1971} extended the Poisson process to account for self-excitation in the system by modeling the intensity conditional on the history of the process $\fil_t=\{t_i: t<t\}$: $\lambda(t|\fil_t) = \lim_{\Delta\to 0} \Delta^{-1} \E{ N_{t+\Delta} - N_t \vert \fil_t}$. In the standard univariate Hawkes model, this intensity reads
\begin{equation} \label{eq:hawkes_def}
	\lambda(t|\fil_t) = \mu +\int_{-\infty}^t \phi(t-s) \diff N_s = \mu + \sum_{t_i<t} \phi(t-t_i),
\end{equation} 
where the first part ($\mu$) gives the so-called \emph{baseline intensity} of the classical exogenous Poisson process and the second one describes the explicit endogenous impact of the past events on the future. Here $\phi(t)$, called the \emph{memory kernel}, is a non-negative function that specifies how past events contribute to the generation probability of future events and thus $\phi(t)$ controls the amplitude of the feedback mechanism. 

The linear structure of the intensity~\eqref{eq:hawkes_def} allows one to map the Hawkes process exactly onto a cluster process~\citep{hawkes_oakes1974}, where the process consists of a superposition of random clusters, each of which starts with a single \emph{immigrant}, generated from a homogeneous Poisson process with intensity $\mu$. In turn the immigrants generate their offsprings according to an inhomogeneous Poisson process with intensity $\phi(t)$, and these next-order events generate their own offspring with the same mechanism. Such construction can be well-described using the theory of branching processes~\citep{VereJones2008_vol2}, and from a practical perspective such representation allows to perform efficient numerical simulation of the Hawkes process~\citep{moller2005}. The branching context also provides an intuitive constraint on the kernel $\phi(t)$ that ensures the stability and stationarity of the system. The average number of offsprings generated by a single event is $n=\int_0^\infty\phi(t)$, called \emph{branching ratio}. For the process to be stable, this quantity has to be smaller than 1 ($n<1$). The condition $n=1$ corresponds to a tipping point and the system with $n>1$ exhibits exploding dynamics with total population increasing to infinity with probability one.

In \citet{Rambaldi2015} authors have further extended the Hawkes model.
Instead of insisting that all immigrants arrive at random times following a homogeneous Poisson process, authors have proposed that some external events could significantly affect the evolution of the system. Such external events can locally give rise to many more immigrants than under the sole baseline intensity. 
This was modeled via a separate term in the intensity expression:
\begin{equation} \label{eq:news_model}
	\lambda(t)= \mu +  \sum_{j=1}^M \phi^j_S (t- z_j)  + \sum_{t_i < t} \phi(t-t_i) 
\end{equation}
Here the second term describes the impact of $M$ exogenous events arriving at times $z_j$, that are assumed to be deterministic and known. These special exogenous events increase the rate of immigrants arrival via the memory kernels $\phi^j_S(t)$. The system further reacts in a regular way via the kernel $\phi(t)$ thus amplifying the effect of the external events.

As a matter of fact, these special exogenous events and their clusters introduce IBs into the system. For example, within the modeling of socio-economic systems, such intensity bursts could correspond to the reaction of the system to major events, such as elections, referendums and other political events, regulatory changes or even announcements of news related to a particular sector or company.

The expected number of new immigrants generated by a single exogenous IB is
\begin{equation}\label{eq:fertility}
	f_j = \int_{0}^{+\infty} \phi^j_S (s) \diff s,
\end{equation}
and we will refer to this quantity as the {\it fertility} of the $j$-th IB. In contrast to the branching ratio $n$ of the whole system, there are no restrictions on this parameter, except that the integral in~\eqref{eq:fertility} should converge. Moreover, we will be interested in cases when the exogenous event represent a ``burst'', and thus its fertility is much larger than of a regular event: $f_j\gg1$.  The immigrants directly triggered by the IB will produce their own offsprings according to the memory kernel $\phi(t)$ and thus the expected total number of events in each exogenous cluster is given by
\begin{equation}\label{eq:shock_cluster_size}
	\Nshock_j = f_j + f_j\frac{n}{1-n} = \frac{f_j}{1-n} 
\end{equation}

Finally, in order to fully specify the model, we need to define the functional form of memory kernels. There is no unique specification and the shape of the memory kernel will vary from one application to another. Within the domain of modeling high-frequency financial data, which is the application discussed in this paper, the question whether the kernel $\phi(t)$ should be short- or long-memory is rather controversial. Some works show empirical support for the power law decaying functions (see e.g.~\citet{bouchaud_hawkes,BacryMuzy2014_secondorder}), while other works suggest use of sum of a few exponential functions~\citep{LallouacheChallet2014,Martins:2016ta} and point out dangers of using long-memory kernels on non-stationary data~\citep{Filimonov:2015fm}.

For our purposes we will follow the original work of \citet{Rambaldi2015} and adopt an approximation of the power-law kernel via a sum of exponentials that was originally suggested in~\citep{bouchaud_hawkes}:
\begin{equation}\label{eq:bouchaud_kernel}
	\phi(t)= \frac{n}{Z} \left\{ \sum_{k=0}^{K-1} (\tau_0 m^k)^{-p} e^{-\frac{t}{\tau_0 m^k} }-S e^{ -\frac{tm}{\tau_0} } \right\}.
\end{equation}
It approximates a power law decay with exponent $p$ and it also features an exponential cutoff at short time scales. The parameter $\tau_0$ controls the position of the maximum and $n$ is the branching ratio. Further, $m=5$, $K=15$ and values of $S,Z$ are fixed such that $\phi(0)=0$ and $\int_0^\infty \phi(s) \diff s = n$. This kernel specification has been shown (\citet{Rambaldi2015}, \citet{bouchaud_hawkes}) to be well suited to model financial data. Moreover, the slow decay of the kernel creates relatively strong point clustering and long range dependencies between counts of the process. This allows us to test our procedure in a somewhat harsher environment, as it makes more difficult to separate genuine external IBs from endogenously generated bursts. 

Finally, for the memory kernel $\phi_S(t)$ of the IB we will follow~\cite{Rambaldi2015} and adopt the exponential specification:
\begin{equation}
\label{eq:exokerexp}
\phi_S (t) =\alpha e^{-\frac{(t-z)}{\tau}}\mathbbm{1}_{t>z} 
\end{equation}
where $\alpha>0$ and $\tau>0$ are parameters. In this case we have $f=\alpha \tau$. This specification reflects the idea that an external event has a very strong immediate impact and also some time persistence, and finally its effects fades completely. Nevertheless, in other domains different specification of $\phi_S$ could be more appropriate, and our procedure could be easily adapted to them.

%================================================================================ 
%================================================================================ 
\section{Identification of the intensity bursts}\label{sec:identification}

In the original work of~\citet{Rambaldi2015}, the times of the IBs $\{z_j\}$ as well as their number $M$ were assumed to be known. Here we assume they are unknown and we extend the framework to define a rigorous procedure for the detection of both the number and the timestamps of the exogenous shocks.

\subsection{Identification of a single IB}

For clarity, let us start with the case where there is at most one IB, but we do not know its occurrence time. In this case our model reads 
\begin{equation}\label{eq:model}
	\lambda(t) = \mu + \sum_{t_i < t} \phi(t-t_i) + \phi_S (t-z) 
\end{equation}
and now $z$ is a parameter of the model. 

In general, once the Hawkes model is specified, we can apply a range of methods in order to calibrate the model on the observed data and estimate its parameters. Such methods include maximum likelihood estimation~\citep{Rubin:1972ii,Ogata1978,Ozaki1979} and method of moments~\citep{DaFonseca:2014hc}. A range of non-parametric tools~\citep{bacry2012non, BacryMuzy2014_secondorder, lewis2011nonparametric, kirchner2015estimation} are also available. These are however mostly limited to the classical Hawkes specification \eqref{eq:hawkes_def}. Within the parametric estimators the Maximum Likelihood Estimator is considered to be the de-facto standard for the Hawkes process family. Given the functional form of the intensity~\eqref{eq:model} and the observations $\fil_T$ in the interval $[0,T]$ one can maximize the log-likelihood
\begin{equation}\label{eq:loglike}
	\log \like (\theta|\fil_t)= -\int_0^T \lambda(s) \diff s + \int_0^T \log\lambda(s) \diff N_s
\end{equation} 
in order to obtain the vector of parameters $\theta$, which includes the time $z$ of the IB and parameters of the kernel $\phi_S$ that defines the fertility $f$.
However, a-priori we do not know if the realization contains any IBs. Thus, we need to decide whether the identified burst in activity is indeed genuine or could be attributed to the endogenous mechanism alone. For this we will test if the extension of the model from a simple Hawkes process~\eqref{eq:hawkes_def} to a model with a shock~\eqref{eq:model} improves the description of the data. A natural approach is to compare the likelihood of model~\eqref{eq:model} evaluated at the optimal parameter with the corresponding best fit from a Hawkes model without IBs. 

\citet{Ogata1978} proved under certain regularity assumptions that the maximum likelihood estimate for a simple, stationary, univariate point process is consistent and asymptotically normal as the sample size tends to infinity. Moreover, he also established that the likelihood ratio test of a simple null hypothesis possesses the standard $\chi^2$ distribution. Therefore, as suggested also in~\cite{Gresnigt2015}, likelihood tests such as the Likelihood Ratio (LR) test and the Lagrange Multipliers (LM) test can in general be used to discriminate between different Hawkes specification. 

However, our case presents some difficulties that make such tests not readily applicable. Our situation is akin to that found in regime shift problems when the change point in unknown. Moreover, some of the parameters are not identified under the null hypothesis that no IB is present. Indeed, the model with the IB term described by~\eqref{eq:exokerexp} reduces to the null model when $\alpha=0$. Hence, $\tau$ and $z$ are not identified under the null. One can also argue that the reduction to the null is not unique, since the effect of the IB is removed also for $z\ge T$, or $\tau\to0$, or $z<0$ and $\tau\to \infty$. As discussed in \citep{andrews1993tests, davies1977hypothesis, davies1987hypothesis, Hansen:1996bd} the standard asymptotic theory of LR does not apply in these cases and other approaches have to be taken. 

One stream of literature deals with this problem through the so called ``sup'' class of tests (see~\citet{Lange:2009fe} for a survey), which rely on simulations of the LR distribution under the null to compute the appropriate p-values. 

Another approach \citep{Wong:2001hz} uses information criteria for model selection. The most widely employed are Akaike information criterion
\begin{equation}
	\AIC = 2k- 2\log\like
\end{equation} 
where $k$ denotes the number of estimated parameters, and the Schwartz (or Bayesian) information criterion (BIC) 
\begin{equation}
	\BIC = k \log N - 2\log \like
\end{equation}
which penalizes more heavily extra parameters if, as in our case, the sample size $N>7$ (i.e.\ $\log N>2$). In this paper we follow this second approach, and will document that the BIC performs well in our test (see Section~\ref{sec:model_validation}).

We can summarize our procedure for the identification of a single IB as follows:
\begin{enumerate}
\item the null model (Hawkes model without any IB term~\eqref{eq:hawkes_def}) is estimated using the Maximum Likelihood Estimator;
\item the alternative (extended) model~\eqref{eq:model} is estimated with the constraint of $z\in [0,T]$;
\item the score $\Delta \mbox{BIC} = \mbox{BIC}_{1} -\mbox{BIC}_0$ is evaluated. If $\Delta \mbox{BIC} <0$ then the null model is rejected and we accept the extended model. Otherwise we retain the null model of no exogenous bursts.
\end{enumerate}
Here we have denoted with $\mbox{BIC}_{M}$ the score of the model with $M$ IBs. In case of exponential memory kernel of IB~\eqref{eq:exokerexp} the difference in numbers of parameters between the null and the alternative model is 3 and thus $\Delta \mbox{BIC} = 3\log N - 2(\log\like_1-\log\like_0)$.
We refer to \ref{app:like} for more details on the likelihood optimization. 
%================================================================================ 
\subsection{Identification of multiple IBs: pre-identification}

We now discuss the extension of our procedure to the case of more than one IB in the window $[0,T]$, i.e. the model \eqref{eq:news_model}, where $z_j$ are parameters of the model and also the total number of IBs $M$ has to be determined. Because of the numerical complexity of the problem, often noisy data, and the curse of dimensionality, straightforward estimation of the full model~\eqref{eq:news_model} leads to an optimization of a multi-variate cost-function with multiple local extrema. In order to efficiently estimate the parameters $z_j$ in \eqref{eq:model} it is useful to restrict the search space for each $z_j$ to some interval $[z_1^j,z_2^j]\subset [0,T]$. Doing so significantly improves the convergence and reduces the number of local minima. Moreover, it allows us to focus on one IB at a time and thus to increase sequentially the size of the model. 

The problem we address in this section is thus how to efficiently and reliably reduce the search space for $z$ via pre-identification. Later we will present the complete procedure.

We adapt the method, originally proposed by~\cite{Almgren:2012wq}, for detection of price jumps in high-frequency data to a point process setting. Following his work we define the exponential averaging functions $u_L(t;\almgtime)$ and $u_R(t;\almgtime)$ as 
\begin{equation}\label{eq:ul}
u_L(t;\almgtime) = \frac{1}{\almgtime} \int_{-\infty}^{t}  e^{-\frac{(t-s)}{\almgtime} }\diff N_s = \frac{1}{\almgtime} \sum_{t_j<t} e^{-\frac{(t-t_j)}{\almgtime}}
\end{equation}
\begin{equation}\label{eq:ur}
u_R(t;\almgtime) = \frac{1}{\almgtime}\int_{t}^{+\infty}  e^{-\frac{(s-t)}{\almgtime} }\diff N_s = \frac{1}{\almgtime} \sum_{t_j>t} e^{-\frac{(t_j-t)}{\almgtime}},
\end{equation}
where $\almgtime$ is a parameter of the method.

\begin{figure}[htb]
\centering
\includegraphics[width=0.9\textwidth]{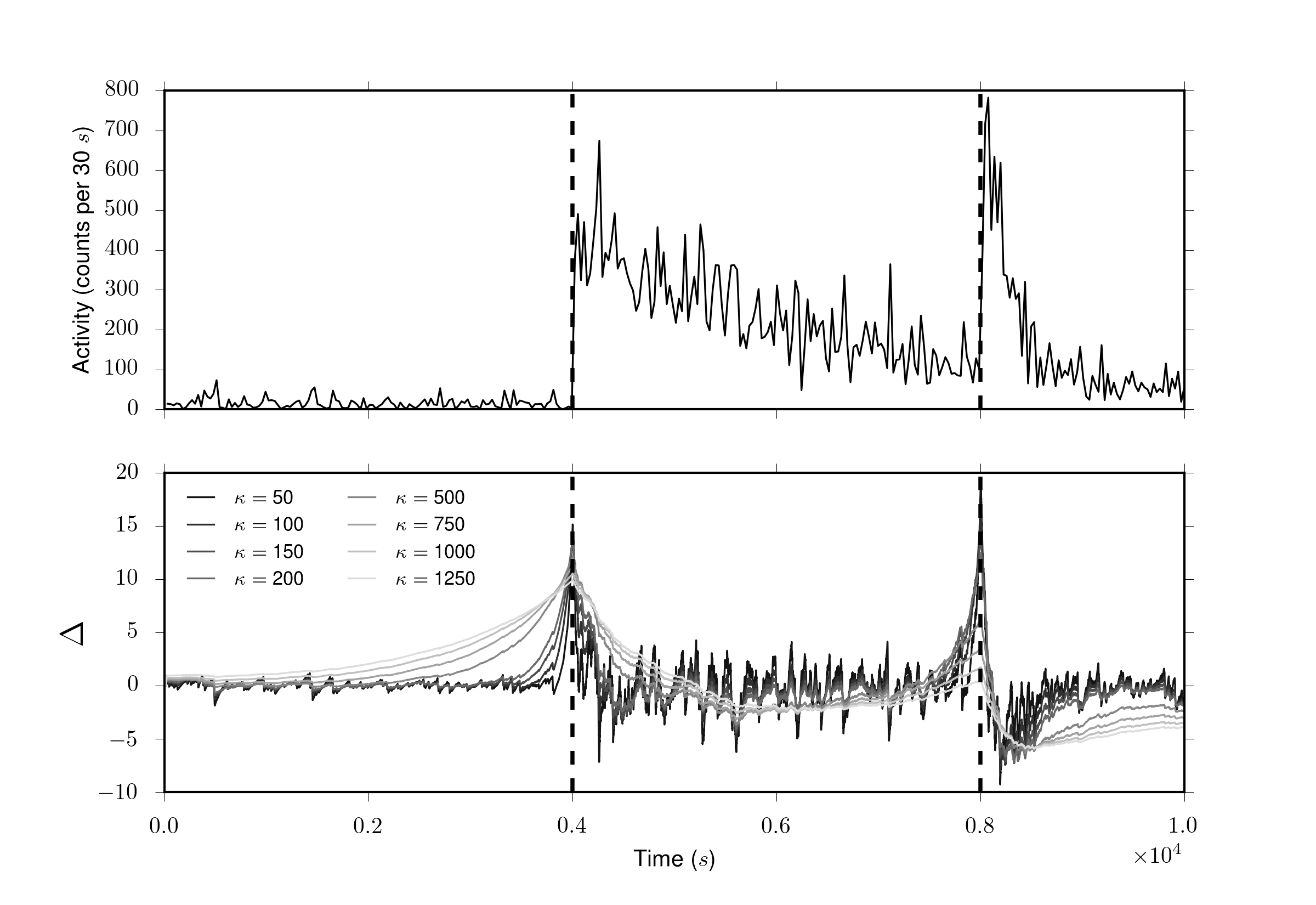}
\caption{Simulation of a Hawkes process with two IBs (top panel) and the corresponding functions $\Delta(t;\almgtime)$ for different choices of $\almgtime$ (bottom panel).}
\label{fig:delta_example}
\end{figure}

The functions $u_L(t;\almgtime)$ and $u_R(t;\almgtime)$ provide a local estimate of the rate of the process. The first one considers only the past, while the second one considers only the future\footnote{In real time analysis it is impossible to calculate $u_R(t;\almgtime)$. However, here we are interested in ex-post analysis of the historic data, so this limitation does not pose any problems.}. The strong sudden jump in activity (IB) at time $t$ will dramatically increase $u_R(t;\almgtime)$ and will not contribute to $u_L(t;\almgtime)$, thus we consider the difference
\begin{equation}
	\Delta(t;\almgtime) = u_R(t;\almgtime)-u_L(t;\almgtime) 
\end{equation}
in order to identify the times of jumps. An example is provided in Figure \ref{fig:delta_example}, where the function $\Delta(t,\almgtime)$ is shown for different values of $\almgtime$. The choice of $\almgtime$ will be discussed in Section~\ref{sec:pre_id_valid}. High values of $\Delta$ correspond to abrupt changes in the intensity of the process and thus provide a way to identify probable IB locations\footnote{For this purpose it is particularly efficient to evaluate $\Delta(t;\almgtime)$ only at the event times. In fact, the recursive relations
$$\almgtime\, u_L(t_i) = e^{-\frac{t_i-t_{i-1}}{\almgtime}} \left(1 +  \almgtime\, u_L(t_{i-1})\right)$$
$$\almgtime\, u_R(t_i) = e^{-\frac{t_{i+1}-t_{i}}{\almgtime}} \left(1 +  \almgtime\, u_R(t_{i+1})\right)$$
hold.}.
  
As candidates for the IB times $\bar{z}_j$ we select the local maxima of $\Delta(t;\almgtime)$ over the set of the event times $\lbrace t_i \rbrace$. Moreover, we rank them from the most to the least relevant one, according to the value of $\Delta$ at the maximum. We start with the global maximum
\begin{equation}\label{z1}
	\bar{z}_1 = \argmax_{t \in \lbrace t_i \rbrace } \Delta(t; \almgtime),
\end{equation}
which defines a window $W_1 = [\bar{z}_1 -\frac{w}{2},  \bar{z}_1 +\frac{w}{2}]$ of size $w$ around $\bar{z}_1$ where the optimal value of $z_1$ will be searched using the maximum likelihood estimator. We furthermore exclude all events which are closer to $\bar{z}_1$ than $w$ and look for next local extremum:
\begin{equation}\label{z2}
	\bar{z}_2 = \argmax_{t \in \lbrace t_i: |t_i-\bar{z}_1|>w \rbrace} \Delta(t; \almgtime),
\end{equation}
which defines another window $W_2 = [\bar{z}_2 -\frac{w}{2},  \bar{z}_2 +\frac{w}{2}]$.
The exclusion is important in order to select distinct maxima of the function $\Delta$: as it is seen from Figure~\ref{fig:delta_example}, the function $\Delta$ is rather persistent and if this precaution is not taken, then the second-to-best maximum will be most likely selected next to the global maximum. 
The procedure is repeated as many times as necessary to find the candidate times
\begin{equation}\label{zi}
	\bar{z}_{k} = \argmax_{t \in \lbrace t_i:|t_i-\bar{z}_1|>w, |t_i-\bar{z}_2|>w,\dots,|t_i-\bar{z}_{k-1}|>w \rbrace} \Delta(t; \almgtime)
\end{equation}
and corresponding windows $W_k = [\bar{z}_k -\frac{w}{2},  \bar{z}_k +\frac{w}{2}]$

%================================================================================ 
\subsection{Identification of multiple IBs: estimation} 
\label{sec:complete_model}

Once a ranking of the windows $W_j$ has been determined, we apply an iterative procedure to determine the optimal parameters of the model $\eqref{eq:news_model}$ as well as the total number $M$ of IBs. First, we estimate the model with $M=0$, that is a standard Hawkes model. Then, we estimate the model with $M=1$ and $z_1 \in W_1$. At this point we use the BIC (or another model selection criterion) to decide whether to accept or reject the enlarged model. If the IB is not accepted, the procedure stops and the null model with $M=0$ is selected. Otherwise, the model is extended to the case $M=2$ and an additional IB term is added. The optimal value of $z_2$ is estimated within the window $W_2$, while keeping the value of $z_1$ fixed, and all the other parameters are reoptimized. The information criterion is again used to compare the penalized likelihood of the $M=2$ model to that of the case $M=1$: then the new extended version is either accepted or rejected. The iterative procedure stops when the addition of the $M_{k+1}$ IB does not improve the likelihood of the model significantly over the model $M_k$.

We can summarize our complete procedure as follows:
\begin{enumerate}
\item Given a realization $\lbrace t_1, t_2, \dots, t_N\rbrace$, we select values for $\almgtime$ and $w$.
\item We determine a ranking of the candidate IB locations $W_1, W_2, \dots$.
\item We estimate a standard Hawkes model on the data ($M=0$)
\item We add to the model one IB at a time
\begin{itemize}
\item If $\text{BIC}_{M=k+1} < \text{BIC}_{M=k}$ the candidate IB $z_{k+1}$ is added to the model and the next one is examined,
\item Else if  $\text{BIC}_{M=k+1} > \text{BIC}_{M=k}$ the procedure stops and we consider $M=k$ as the final number of IBs.
\end{itemize}
\end{enumerate}

We note that this procedure relies on the ranking provided by the pre-identification procedure. Stopping criteria that rely less on the pre-identification procedure can also be considered. For example, instead of stopping at the first failure, additional IBs can be considered before exiting the loop. Again, we refer to \ref{app:like} for details on the optimization procedure.

%================================================================================ 
%================================================================================ 
\section{Numerical simulations and model validation}
\label{sec:model_validation}

In order to validate our procedure, we performed extensive numerical simulations. First, we examined the rate of false positives produced by our procedure. Then we tested its statistical power. Finally, we studied how discrimination works in the case where two IBs are present in the window. Since even in the case of the standard Hawkes process~\eqref{eq:hawkes_def}, bias and variance of the estimations of the parameters depend on the values of parameters themselves, we test several combinations of the parameters on various sample sizes in order to get a broad view on the efficiency of the detection procedure.

\subsection{Absence of exogenous shocks}
\label{sec:no_shock_valid}

In order to study the rate of false positive detections we performed Monte-Carlo simulations of a Hawkes process~\eqref{eq:hawkes_def} with the endogenous kernel~\eqref{eq:bouchaud_kernel} and without any IB. The parameters are $p=2.0$, $\tau_0 = 0.1$,  and $n=\lbrace0.3,0.5,0.7,0.9\rbrace$. We simulate 1000 realizations for each parameter set. Because the power of the statistical test strongly depends on the sample size, in our simulations we keep the expected sample size $\E{N}=\mu T/(1-n)$ constant by fixing simulation horizon $T=3,600$ and varying the baseline intensity $\mu$.
We perform tests with sample sizes of approximately 1,000, 2,000, 5,000, and 10,000 events. 

\begin{figure}[htb]
\centering
\includegraphics[width=0.4\textwidth]{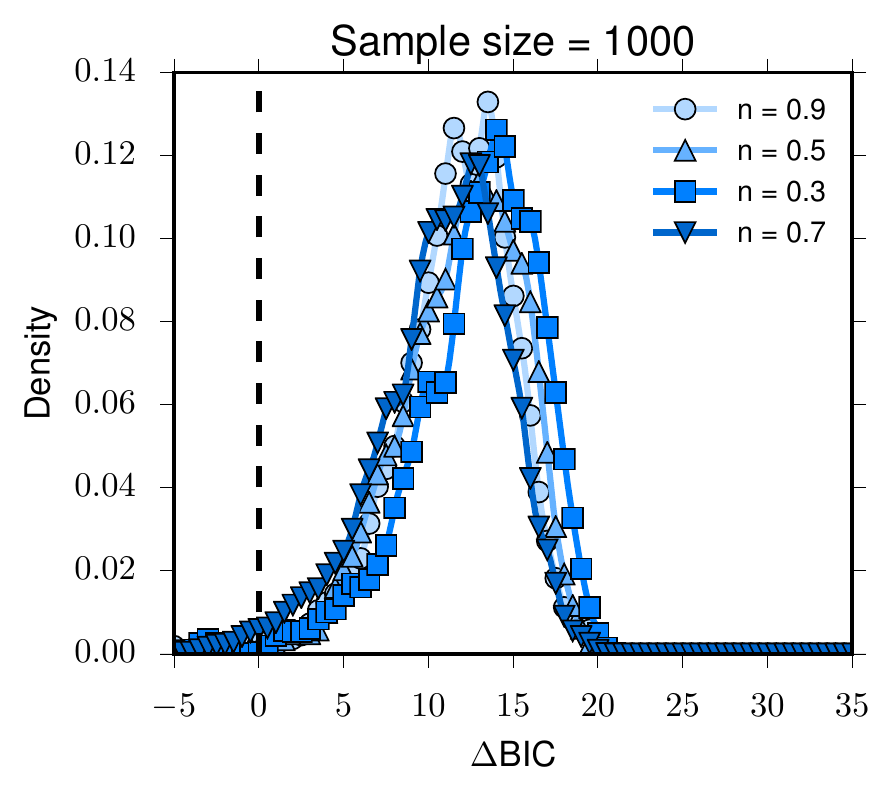}
\includegraphics[width=0.4\textwidth]{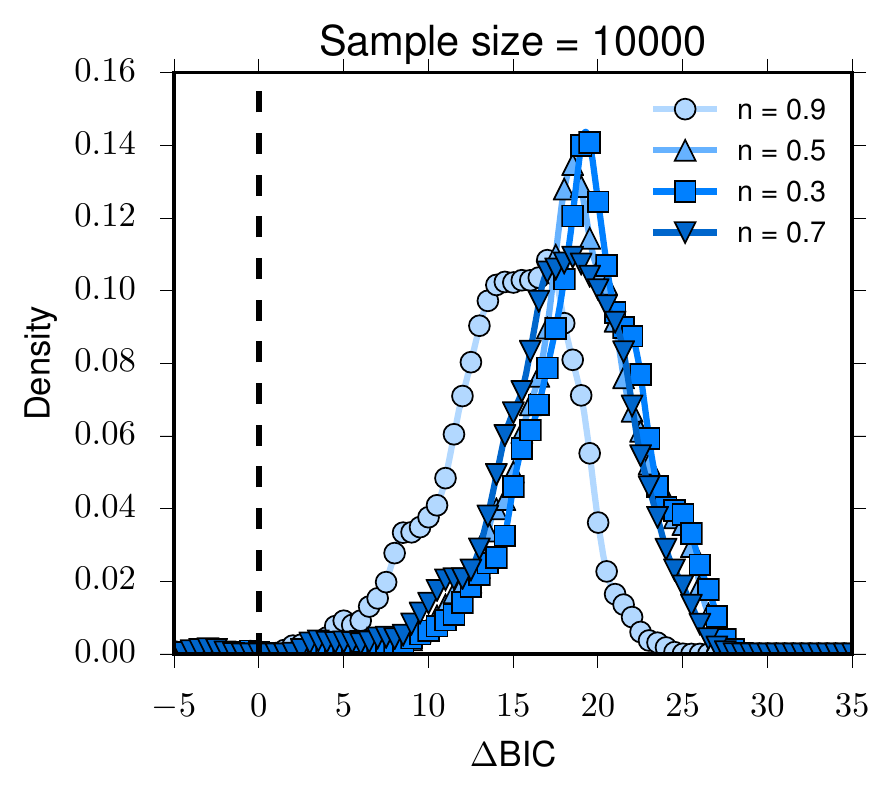}
\caption{Distribution of the differences $\Delta$BIC  between the null model where no IB term is present and the complete model with a single IB estimated on simulations where no IB is present. For differences smaller than zero the null model is rejected.}
\label{fig:aic_bic_nonews}
\end{figure}

In each simulation we apply our procedure and check whether or not the extended model \eqref{eq:model} is preferred over the standard Hawkes model. In general we found that AIC performs much worse than BIC, especially for high values of $n$. The rates of false positives, expressed in percent, resulting from the BIC are summarized in Table \ref{tab:FP}. As it is seen from the table, even for moderate sample sizes (1000 events) and very strong self-excitation of the underlying process ($n\sim0.9$) the rate of false positives is around 1\%, and in general the specificity of the method is close to 100\%. This is further illustrated by the Figure \ref{fig:aic_bic_nonews} that presents the distribution of the observed scores $\Delta$BIC.

\begin{table}
\centering
\input{./Tables/FPR_noshocks.tex}

\caption{Percentage of false positives using Bayesian Information Criterion when the simulated model has no IBs. All values expressed in percent (\%).}
\label{tab:FP}
\end{table}

%================================================================================ 
\subsection{Single IB with approximately known location}
\label{sec:ss_0}

We now consider numerical simulations when one IB is present. Here we present results of simulations with parameters $p=2.0$, $\tau_0=0.1$, $n=\lbrace0.3,0.5,0.7,0.9\rbrace$. The simulation window is $[0,T]$ where $T=3,600$. An IB is simulated at $z=T/2$, and 35 different combinations of IB parameters ($f, \tau$) are considered (see below for details). As before, we vary the value of $\mu$ in order to keep the expected sample size 
\begin{equation}
	\frac{\mu + f (1-e^{\frac{z-T}{\tau}})}{1-n}T
\end{equation}
approximately constant. We performed two sets of simulations with average sample sizes 5,000 and 10,000 respectively.
Since the primary concern of this experiment is the rate of false negatives resulting from our procedure and our choice of the BIC selection criterion, here we ignore the pre-identification routine and we will address the complete procedure in the next sections. We thus limit the search space for $z$ to a window of size 100 centered around the true position of the IB: $[\frac{T}{2}-50,\frac{T}{2}+50]$. 

We report the results of the test in~\ref{sec:appendix_validation}. Table \ref{tab:one_shock_fixed} presents the percentage of correctly detected IBs for different values of the branching ratio $n$ for the sample size of 5,000. Our results show that there is a region corresponding to low values of the fertility $f$ and high values of the relaxation time $\tau$ of the IB (i.e. small values of $\alpha=f/\tau$), where the addition of the IB term does not contribute significantly to the improvement of the likelihood and therefore the extended model is rejected. 

Hawkes processes naturally produce clusters of events through the self-exciting mechanism. The average size of a cluster is given by $\frac{1}{1-n}$ (see e.g. \cite{moller2005}) and its variance is $\frac{n}{(1-n)^3}$. Hence, if an exogenous IB produces an events cascade of similar size to the ones generated endogenously, then our procedure will not identify it as an IB. This is actually a desired feature since we want to identify IBs that are not compatible with the endogenous dynamics. Moreover, this also suggests that for high values of the branching ratio $n$ it becomes more difficult to distinguish between an endogenously generated burst of events and an exogenous one. 

%================================================================================ 
\subsection{Pre-identification}
\label{sec:pre_id_valid}

The efficiency of the pre-identification algorithm depends on the choice of the ``smoothing parameter'' $\almgtime$. When $\almgtime$ is small, the approximation of the local intensity will be noisy and will result in detection of many local maxima. On the other hand, large values of $\almgtime$ result in averaging out most of the intensity fluctuations and leaving only big IBs to stand out, but potentially missing medium scale ones. 

In order to test the effect of the choice of $\almgtime$, we simulate the model \eqref{eq:model} with different combinations of the parameters $(\alpha, \tau)$ and then apply our pre-identification algorithm with various values of~$\almgtime$. Results are reported in~\ref{sec:appendix_validation}, where Figure \ref{fig:mse_z} presents the Root Mean Squared Error on the detected IB location relative to the average distance between events $\delta={T}/{N}$ as a function of the ratio $\almgtime/\tau$.

As expected, for small values of $\alpha$, when detection of the exogenous IB is difficult, error of the estimation of $z$ could be very large. However, we observe that for relatively high values of $\alpha$ the error drops to the value comparable with inter-event interval $\delta$, and the best performance is attained when $\almgtime$ is of the same order of magnitude of $\tau$. When $\almgtime$ is much larger than $\tau$ the performance deteriorates also for high values of $\alpha$. This suggests that if the typical relaxation time of the IB is known then the best choice is to select $\almgtime$ to be of the same order. Alternative but more complicated way could be to employ the multi-scale analysis for multiple values of $\almgtime$ in a spirit of~\citep{Almgren:2012wq}. However, given that (i) our method is parametric and pre-identification is followed by the maximum likelihood estimation and (ii) due to stochastic nature of the Hawkes process~\eqref{eq:news_model}, the peak of local intensity does not necessarily coincides with the start time $z$ and can be delayed depending on the kernel $\phi(t)$, we consider the multi-scale approach to be an over-complication.

%================================================================================ 
\subsection{Single IB detection with the complete procedure}
\label{sec:ss_1}

We now test our complete procedure in the case where one IB is present. We follow the same procedure as in Section \ref{sec:ss_0}. However, now we do not impose the search interval for $z$, applying the pre-identification method instead. We set $\almgtime=100$ and we set the search window size in~\eqref{z1}--\eqref{zi} to  $w=300$, i.e. while doing maximum likelihood estimation, we perform a constrained search for $z_j$ in windows $W_j=[ \bar{z}-150, \bar{z}+150]$. We consider the actual IB to be correctly identified by our procedure when one of the detected IBs is within a tolerance of $60$ time units from its occurrence time. This includes also the cases where the actual IB is not the first one to be detected.

Table \ref{tab:one_shock_complete} reports the percentage of correctly classified IBs (true positives) for the different combinations of $n$, $f=\alpha\tau$, and $\tau$. The performance of the method is shown to be comparable with the case where the correct search interval for $z$ was known in advance (compare with Table \ref{tab:one_shock_fixed}), indicating that the pre-identification algorithm works fairly well in most cases. The lower rate of correct detections are found unsurprisingly for high values of $\tau$ and small values of $f$, i.e. small values of $\alpha$. In these cases, the immediate shock to the intensity by the IB is smaller and the cumulative effect of the IB is more diluted over time, which makes it harder to localize it.

\begin{table}[htb]
\small
\centering
\input{./Tables/one_shock_size_5000_n_0.3.tex}
\input{./Tables/one_shock_size_5000_n_0.5.tex}
\vspace{5pt}

\input{./Tables/one_shock_size_5000_n_0.7.tex}
\input{./Tables/one_shock_size_5000_n_0.9.tex}
\caption{Percentage of correctly classified IBs for different combinations of the true IB parameters $(\alpha, \tau)$ expressed in terms of $f=\alpha \tau$ and $\tau$. The results refer to a sample size of roughly 5,000 events.}
\label{tab:one_shock_complete}
\end{table}

Further we need to explore the rate of false positives, i.e. the cases when more than one IB is detected in these simulations. 
Table~\ref{tab:fp_one} reports the number and the percentage of false positive identifications as a function of the branching ratio $n$. As it is seen, the number of cases where more than one IB is detected is however limited and concerns mainly the simulation of high values of $n$, namely $n=0.7$ and especially $n=0.9$. This is expected, since in these cases fluctuations of the intensity of the background Hawkes process are very strong and the probability of spontaneously generating a cluster which might me classified as IB is high. So reliable detection here requires sufficiently long time series, which ensures the efficient estimation of parameters of the Hawkes process when $n$ is high.

\begin{table}
\small
\centering
\begin{tabular}{lcccc}
\toprule
& \multicolumn{4}{c}{Sample size = 5000}\\
\midrule
& $n=0.3$ & $n=0.5$ & $n=0.7$ & $n=0.9$ \\
\midrule
Total FP & 1 & 3 & 12 & 63\\ 
Worst case incidence (\%) & 0 & 0 & 2 & 18 \\
\midrule
& \multicolumn{4}{c}{Sample size = 10000}\\
\midrule
Total FP & 0 & 0 & 9 & 37\\ 
Worst case incidence (\%) & 0 & 0 & 4 & 14 \\
\bottomrule
\end{tabular}
\caption{Total number of false positives and percentage of false positives in the worst case resulting from our procedure applied on simulations where one IB is present. The worst case for a given value of $n$ is the single combination $(f,\tau)$ where the most false positives are recorded.}
\label{tab:fp_one}
\end{table}

Next, we examine the error on the estimation of the IB starting time $z$. In Table \ref{tab:err_z_one} we report the ratio between the root mean squared error $ RMSE[(z - \hat{z})^2]$ and the average inter event time $\delta = T/N$. As before, the results are much better when the parameter $\alpha=f/\tau$ is large, while the performance deteriorates for small $\alpha$ and comparatively high $\tau$ albeit still remaining satisfactory.

\begin{table}[htb]
\small
\centering
\input{./Tables/err_z_size_5000_n_0.3.tex}
\input{./Tables/err_z_size_5000_n_0.5.tex}
\vspace{5pt}

\input{./Tables/err_z_size_5000_n_0.7.tex}
\input{./Tables/err_z_size_5000_n_0.9.tex}
\caption{Ratio between the root mean squared error on the IB location parameter $z$ and the average inter-event time $T/N$. The results shown are for a sample size of roughly 5,000 events. The error is computed only for cases where the IB is detected at least on 50 out of 100 simulations. }
\label{tab:err_z_one}
\end{table}

In our procedure, we use the pre-identification algorithm discussed above to come up with a first guess of the IB location $z_g$. It is worth comparing this initial guess with the final estimate $\hat{z}$ provided by our procedure. We consider the differences:
$$ \frac{|z_g-z|-|\hat{z}-z|}{\delta}$$
where $z$ represents the true IB location, for the correctly classified IB. We observe that in 91\% of the cases the final estimate is not worse than the pre-estimaed guess (i.e. as close or closer to the true value). Moreover, when our method improves on the pre-identification method, the improvement is on average more than twice the average error when our method does not improve.
In other words, the maximum likelihood optimization of the location parameter $z$ improves significantly the estimation, even in such cases when the exogenous kernel~\eqref{eq:exokerexp} is strictly decaying and has maximum at (which leads to the expected peak of the intensity to be at $t=z$). In more complicated cases, such as when the peak of the reaction is delayed, the optimization step becomes essential.  

In~\ref{sec:appendix_parameters}, Tables \ref{tab:mse_alpha_5000}, \ref{tab:mse_alpha_10000}, \ref{tab:mse_tau_5000}, and \ref{tab:mse_tau_10000} report the relative mean squared errors on the IB parameters. For IBs in the detectable region the MSE is typically of the order of 10\% for considered sample sizes. 

Finally, it is worth mentioning that our procedure can be useful even in cases when the detection of the IB is not the main objective, but one is interested in the parameters of the base Hawkes model, such as the branching ratio $n$. It is known that the presence of non-stationarities and bursts dramatically inflate the branching ratio estimated under the assumption of stationarity (i.e. within the model~\eqref{eq:hawkes_def}). This problem has been discussed already in \cite{Filimonov:2015fm} and \cite{Rambaldi2015} and here we illustrate it with the Table~\ref{tab:n_vs_n} where we compare the values of $n$ obtained from the base Hawkes model and from the best model selected by our procedure when one IB is present. Indeed  in case of an intensity burst, the base model overestimates the branching ratio up to the critical level of 1 even for small values of the actual branching ratio ($n=0.3$). In contrast, properly accounting for such exogenous event via the extended model~\eqref{eq:model} allows us to recover the correct value in all cases.

%================================================================================ 
\subsection{Iterative procedure of IB detection}
\label{sec:ts_valid}

In order to assess the performance of our procedure in the case when multiple IBs are present, we examine eight scenarios where two IBs occur in the same $T=3,600$ time window. Specifically, we consider two kinds of IBs, one that we name ``Small'' (S) with parameters $(\alpha=1.0, \tau=350)$ and another that we name ``Large'' (L) with parameters $(\alpha=1.5, \tau=700)$. The distance between the IBs is also relevant, thus we test separately a configuration where the IBs are close (C), namely $z_2-z_1 =350$ units of time apart, and a second one where the IBs are far (F) apart, namely $z_2-z_1 =1,400$ units. These eight scenarios are summarized in Table~\ref{tab:8cases}.

We simulate 100 realizations for each scenario, for four different values of the endogenous parameter $n$, namely $n=(0.3, 0.5, 0.7, 0.9)$; other endogenous parameters are fixed as in the other experiments to $\tau_0=0.1$ and $p=2.0$ and parameter $\mu$ is adjusted so to keep the expected sample size equal to 10,000 events. Table \ref{tab:two_shocks} presents the result of this test for the case $n=0.7$, while the other cases are reported in Table \ref{tab:two_shocks_app}. 

Our results show that for intermediate values of $n$, our procedure attains very good results. As it is seen from the Table~\ref{tab:two_shocks}, in most of situations rate of true positives exceed 90--95\% with a reasonably low numbers of false positive and false negative detections. The acceptance of more than two IBs in the test (false positive detection) has a maximum value below 10\%  and affecting almost exclusively the $n=0.9$ case. As expected, the probability of false negative increases with the decrease of size of the IB.

The most challenging scenarios for the identification are unsurprisingly those where a small IB follows a bigger one and thus gets overshadowed by the latter. This is especially pronounced when the IBs are close (CLS): in these cases the true positive rate is about 50\% in case of $n=0.7$, and similar in the case when small IB is followed by the small IB (CSS). Further, large values of the branching ratio are more difficult to handle, as we have already noted: when $n$ approaches the critical value of 1, the variance of the cluster size becomes very large, thus making it extremely difficult to differentiate between an endogenously generated burst in intensity and an exogenously generated one.

\begin{table}[tb]
\centering
\begin{tabular}{ccccccc}
\toprule
& $\alpha_1$ & $\tau_1$ & $z_1$ & $\alpha_2$ & $\tau_2$ & $z_2$\\
\midrule
CSS & 1.0 & 350 & 1625 &1.0 & 350 & 1975 \\
CSL &1.0 & 350 & 1625 & 1.5 & 700 & 1975 \\
CLS & 1.5 & 700 & 1625  &1.0 & 350 & 1975 \\
CLL & 1.5 & 700 & 1625  & 1.5 & 700 & 1975 \\
FSS &1.0 & 350 & 1100 &1.0 & 350 & 2500 \\
FSL &1.0 & 350 & 1100 & 1.5 & 700 & 2500 \\
FLS & 1.5 & 700 & 1100 &1.0 & 350 & 2500 \\
FLL & 1.5 & 700 & 1100 & 1.5 & 700 & 2500 \\
\bottomrule
\end{tabular}
\caption{Summary of the eight tested scenarios for cases with two IBs. C stands for close, F for far, L for large, and S for small (see text for details).}
\label{tab:8cases}
\end{table}

The relative mean squared error on the IB parameters is generally under 10\% for $n=(0.3,0.5,0.7)$. Errors on $\tau$ tend to be larger than those on $\alpha$, especially on small IBs  following a large one where error on $\tau$ can be above 60\%. Errors rise significantly for the case of $n=0.9$, where values of more than 30\% on $\alpha$ are not uncommon.

\begin{table}[tbp]
\centering
\input{./Tables/two_shocks2.tex}
\caption{Results of the tests on simulation with two IBs for the $n=0.7$ case. All quantities are expressed in percent. Parameters of simulations are presented in Table~\ref{tab:8cases}. }
\label{tab:two_shocks}
\end{table}

From the above tests we have validated our procedure and confirmed that the method is capable of obtaining meaningful results, provided that the underlying process is not too close to being critical (i.e. $n=\int_0^\infty \phi(t)\diff t$ is sufficiently smaller than one). Naturally the quality of detection and estimation of parameters increases with the increase of amplitude and distance between shocks. 

%================================================================================ 
\subsection{Misspecification of the endogenous kernel}
\label{sec:km}

Finally, we address an important practical aspect of the model error. In real life we do not know a priori what functional dependence of the kernels is the best to describe the dynamics of the system. As it was mentioned above, the discussion about the specification of the Hawkes model~\eqref{eq:hawkes_def} is still open, and for example in financial applications along with approximate power law kernel~\eqref{eq:bouchaud_kernel} a short-memoried exponential alternatives are considered and advocated. The question of the statistical tests for selecting the best candidate as well as the question of non-parametric estimation of the kernel are well beyond the scope of the paper. Here we will address this problem by considering the estimation error in case when the memory kernel is wrongly specified.

For this we simulate the process with a different endogenous kernel (namely --- exponential and double exponential), while performing estimations using the same methodology as before that uses the $\phi(t)$ in the form~\eqref{eq:bouchaud_kernel}. We conduced both experiments where no IBs is present and where one IB is present. Detailed results are presented in~\ref{app:km}. Overall our results are in line with those of Section~\ref{sec:ss_0} and \ref{sec:ss_1}, and show that our procedure is fairly robust with respect to the kernel misspecification: as long as the exact parameters of the underlying Hawkes process is not in the main focus of attention, the exogenous IBs could be identified reliably even when the background process is not well specified.

%================================================================================ 
%================================================================================ 
\section{Application to high-frequency FX data}
\label{sec:fx}

We apply our methodology to study intensity bursts in the spot Foreign Exchange (FX) markets. Currency market is a complex decentralized system of trading platforms and venues, with co-existing Electronic Communication Networks (ECNs), operating in a similar way to regular stock exchanges, and OTC dealing over the terminal chats or voice calls. At the center of the market system there exist two inter-dealer electronic platforms such as Electronic Broking Services (EBS) and Reuters. They are used by major banks and brokers as a source of interbank liquidity and also as a trading platform for large HFT players. Both platforms have a high requirement for the lot size to be at least one million units. 
EBS is the main venue for all the USD, EUR, CHF and JPY crosses, while Reuters is a main platform for trading of all the crosses for Commonwealth currencies and Scandinavian currencies~\citep{Olsen2013}.

%================================================================================ 
\subsection{Descriptive statistics}

We analyze data from the EBS Live data feed for three currency pairs, namely EURUSD, EURJPY and USDJPY. EURUSD is by far the most liquid currency pair in the world with the average daily turnover of 1,289 billions of USD in 2013 according to the BIS triennial survey~\citep{BIS2013}, with the second one being USDJPY with the turnover of 978 billions. EURJPY is also among the most actively traded contracts with the average daily turnover of 147 billions of USD (having only USDGBP, USDAUD, USDCAD and USDCHF before).

Our dataset spans the period from January 1, 2012 to December 18, 2012. FX markets are active 24 hours a day and operate 7 days a week. Activity during weekends is however negligible. 
EBS Live is a premium feed which provides snapshot of the order book every 100 milliseconds (in contrast to a regular feed with quotes every 250 ms). We construct the point process of the events when either of best bid or best ask price changes. The dynamics of such process has been used as a proxy of high frequency volatility and has been modeled with Hawkes processes by several papers (\citet{filimonov2012quantifying,bouchaud_hawkes,Rambaldi2015}).
The limitation of the time resolution of 100ms and absence of records within these sub-intervals might cause bias in the calibration of the point process. The implications of such coarse time resolution on the Hawkes process fitting was discussed in~\citep{LallouacheChallet2014}. To address this issue we use the same randomization procedure of~\citep{filimonov2012quantifying,Rambaldi2015}, namely we subtract from each timestamp a random number uniformly distributed in the interval $[0,0.1)s$.

Our proposed methodology is explicitly designed to detect and model localized ``shock-like'' non-stationarities (IBs) within the Hawkes process framework. However, the methodology is susceptible to other forms of non-stationarity, such as regime shifts or slow changes in base intensity (e.g. day-to-day fluctuations or time-of-day effects). In order to limit the side effects from such dynamics, we limit the window size of the analysis to one hour. Specifically, from the best quote change time series, we extract windows of one hour starting at 00:20:00 UTC of Jan 01 2012. We keep only those windows where a sufficient number of events is present, namely those with at least 2,000 events. Furthermore, since we want to compare results across the three pairs, we retain only windows for which enough events are present in all pairs. This leaves us with 1,551 windows for each currency pair and in the following we will always refer to this subset of the original data. On average, we have 6,394 events per window in the case of the EURUSD pair and 4,169 and 3,187 for EURJPY and USDJPY cases respectively. We apply our complete procedure using $\almgtime = 100$ and $w=300$.

The total number of IBs detected for each currency pair is summarized in Table \ref{tab:n_shocks}. On average we find about 0.3--0.5 IBs per window. We note however that even by selecting small windows on just one hour, we detect a certain number of IBs with very long (much larger than one hour) decay times $\tau$. Most of them are found in the intervals 06:30 - 08:00 and 12:00 - 15:00. The first interval corresponds to the beginning of the European session on the FX market (07:00 London time) and to the opening of European stock markets (08:00 London time). The second one instead is when the American session begins (12:00 London time) and US stock markets open (14:30 London Time)\footnote{Note also that daylight saving times are not synchronized between Europe and North America so there are some weeks of the year when US markets open at 13:30 London time.}. Thus, these shocks appear to capture mostly time-of-day effects, so we have decided to remove them from the analysis. Moreover we disregard bursts with $\tau>5400s$, since the decay time is much longer than the detection time (this amounts to less than 10\% of detected IBs). 

We find that the distributions of the estimated IB parameters $\alpha$ and $\tau$ are very broad, reflecting the diversity of the IBs that we encounter. We also find that these distributions are similar across all currency pairs, and, in particular, the distributions of normalized fertilities $f/N$ are remarkably close among all currency pairs.

\begin{table}[tb]
\centering
\input{./Tables/summary_comb_min2000.tex}
\caption{Total number of IBs detected in each currency pair, average number of IBs per window, fraction of windows with no IB and maximum number of IBs detected in a single window.}
\label{tab:n_shocks}
\end{table}

%================================================================================ 
\subsection{Simultaneous bursts in several markets}

Here we consider IBs that are common between currency pairs, which we defined as events whose estimated starting time $z$ is located within a tolerance window of 60$s$. If we consider simultaneous shocks in all three markets at the same time, they will constitute 15--20\% of all detected IBs. However, this figure increases significantly if we consider only two pairs as shown in Table~\ref{tab:common2}. Since we perform the calibration of the model independently on each FX pair\footnote{A small bias however could be introduced by the fact that we perform our analysis on a set of windows with at least 2,000 events in all three pairs.}, the detection of a large number of common IBs could be considered as a sanity check for the procedure and an indirect evidence that the detected IBs are genuine events. Indeed, in our analysis any two FX rates share one currency in common, hence it is natural to expect that they would react to common drivers. Moreover, all three FX pairs are linked in a triangular fashion and thus are attractive for high-frequency traders who arbitrage away a possible transient mispricing between rates.

\begin{table}[tb]
\centering
\input{./Tables/common2.tex}

\caption{Common IBs between two currency pairs: total number and a fraction of the detected IBs for each pair.}
\label{tab:common2}
\end{table}

\begin{figure}[tb]
\centering
\includegraphics[width=0.5\textwidth]{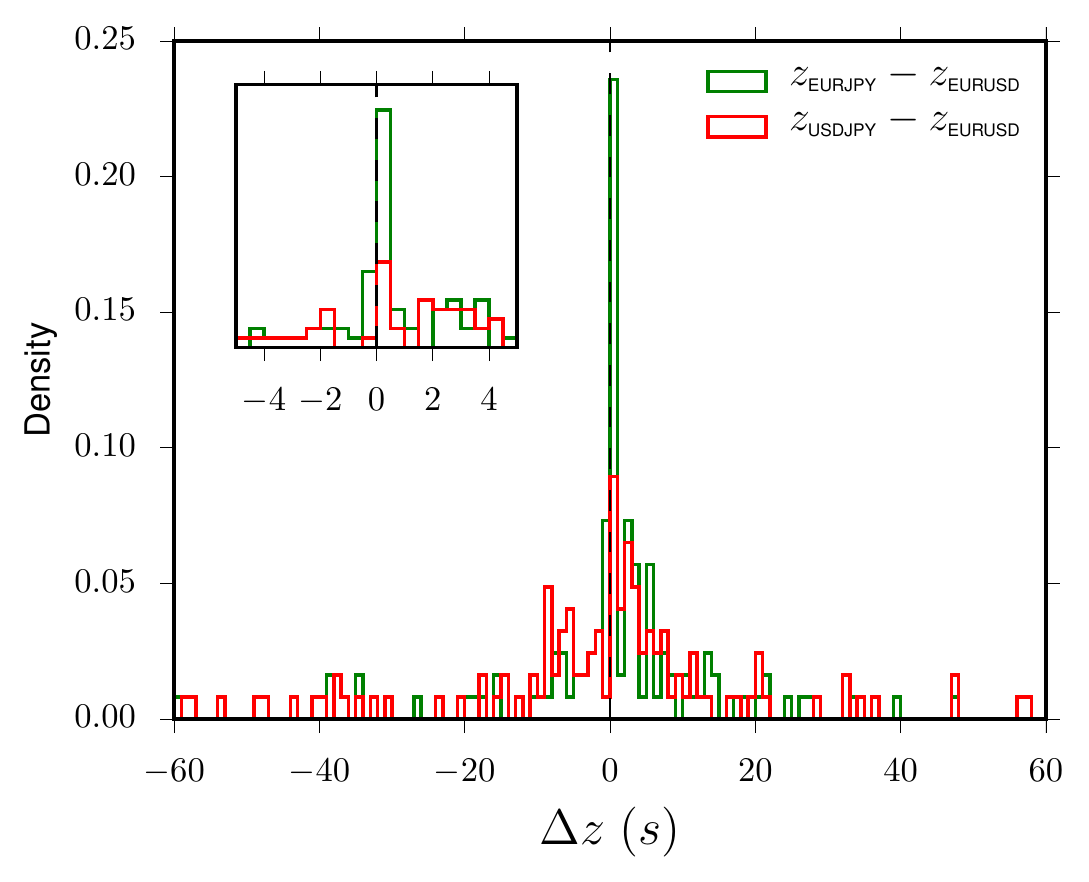}
\caption{Histogram of the differences in detected IB times in the three rates for IBs common to all the three pairs. The time $z_{\mbox{\tiny{EURUSD}}}$ is used as difference. The bin size is 1s in the main plot and 500$ms$ in the inset.}
\label{fig:deltaz3}
\centering
\includegraphics[width=\textwidth]{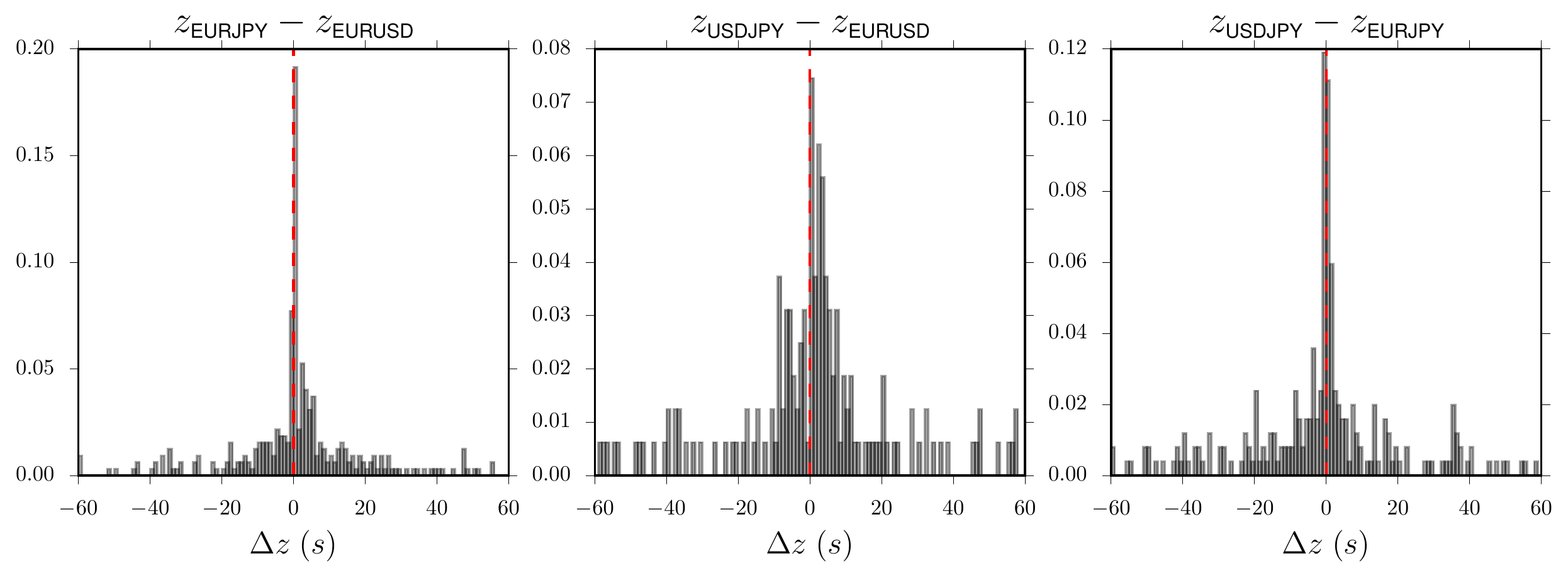}
\caption{Histogram of the differences in detected IB times in two rates for IBs common to two pairs. The bin size is 1$s$. }
\label{fig:deltaz2}
\end{figure}

It is interesting to look at the difference in the estimated initial times $z$ in the case of common IBs. In Figure \ref{fig:deltaz3} we plot the histogram of the time delay of the IBs in EURJPY and USDJPY relative to the shock in EURUSD in case when all three pairs have a common shock. We observe that the distribution is quite broad; however, in both cases the peak of the density is found between 0 and 500$ms$. This suggests that many times EURUSD is the leading pair for an intensity burst, which seems reasonable since it is by far the most liquid and important currency cross. Performing a bootstrap one sided test under the null $z_{xxxyyy}-z_{EURUSD} \le 0$ we obtain t-statisitics (p-values) 0.59 (0.28) and -0.72 (0.77) for EURJPY and USDJPY respectively. If we focus on close matches with absolute difference smaller than 5 seconds, we obtain t-statistics (p-values) 2.07 (0.02) and 2.24 (0.02) respectively. In Figure~\ref{fig:deltaz2} we plot analogous histograms for the common IBs between two pairs. Again, we note that EURUSD appears to lead both the other crosses. Instead, a clear leader does not emerge between EURJPY and USDJPY.

\begin{figure}[tb]
\centering
\includegraphics[width=0.4\textwidth]{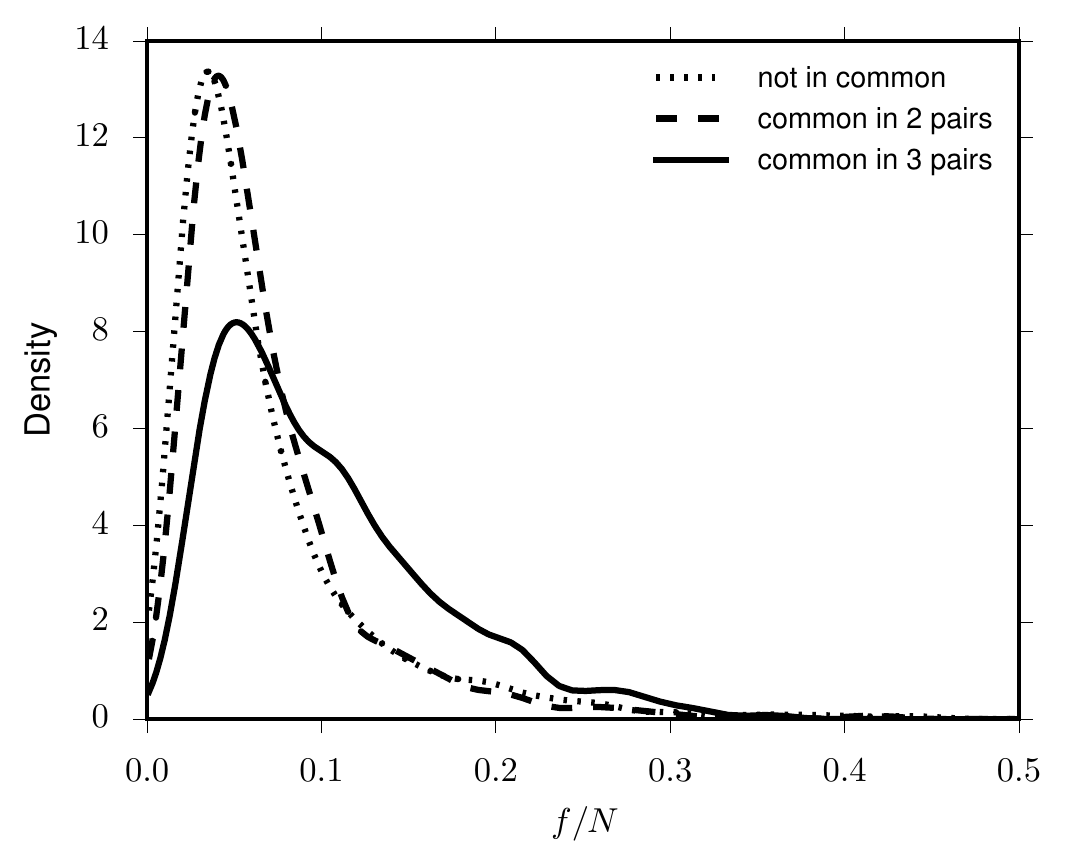}
\caption{Distribution of the ratio $f/N$ for IBs detected simultaneously on all three pairs (solid line) on exactly two pairs (dashed line) and for idiosyncratic IBs (dotted line).}
\label{fig:densityfN}
\end{figure}

Figure~\ref{fig:densityfN} presents the distribution of the normalized fertilities $f/N$ for IBs that are (i) common to all three pairs, (ii) common to only two pairs, and (iii) detected in a single pair only. While the distributions in the cases (i) and (ii) are very close to each other, one can see that the distribution for the third case has much heavier right tail implying that the IBs common to all three pairs have markedly higher fertility. This is confirmed by t-test results in Table \ref{tab:ttest}, while the difference between idiosyncratic IBs and those common to exactly two pairs is not as significant. At the same time Kolmogorov-Smirnov tests are significant at the 1\% level in all cases (including the test for the equality of distributions of (i) and (ii)). 

\begin{table}
\centering
\input{./Tables/ttest_common_noncommon_min2000.tex}

\caption{Results of Welch's t-test for equality of the means and Kolmogorov-Smirnov test for the null hypothesis that the samples come from the same distribution for fertility values of IBs common to all three pairs, to only two pairs and found in a single pair.  }
\label{tab:ttest}
\end{table}

Given the large number of detected IBs, it is natural to ask their possible origin. Two natural explanations are that they are due either to macroeconomic announcements or to price jumps (or both simultaneously). In the following subsections we explore empirically these hypotheses.

%================================================================================ 
\subsection{Intensity bursts and scheduled announcements} 

FX markets are usually very sensitive to macroeconomic announcements such as interest rate decisions, reports on the inflation, GDP and employment data, etc. Most of these announcements are scheduled in advance and market participants prepare for them to account for the uncertainty in the reported numbers. Here we study how the activity responds to such announcements. 

We use a list of scheduled economic announcements retrieved from the website \textit{www.dailyfx.com}. The dataset consists of economic data releases such as inflation, GDP and employment readings, as well as rate decisions communications and press conferences by central banks, for a total of 3,352 separate events. We focus on IBs occurring within $60s$ interval of a planned announcement time in the dataset. Such window size is a good proxy in most of cases, though not always: for example the FOMC rate decision is reported orally by the Chair of the Board of Governors of the Federal Reserve System, and the actual figure of the target interest rate can be pronounced several minutes after the start of the press conference (see for example the reaction time for the fixed income markets in Table 3 of~\citet{Almgren:2012wq}).

In Table \ref{tab:match_news} we report the descriptive statistics of the IBs for which we found a scheduled announcement within a 60$s$ interval. About 10\% of the announcements correspond to a detected IB and conversely about 20\% of the detected IBs are related to an economic announcement. The database of \textit{www.dailyfx.com} also provides a suggestion for the currency most affected by the announcement and an indication of the typical importance of that announcement in a three level scale (Low, Medium, High). Using these indications, we unsurprisingly find that high-importance announcements have a markedly higher detection rate of above 25\%, while figures for low-importance news are around 2\%.

\begin{table}[htb]
\centering
\input{./Tables/summary_news_min2000.tex}

\caption{Number of detected IBs for which there is a scheduled announcement within a 60$s$ interval. The fraction of announcements in the database associated with an IB  is also reported. }
\label{tab:match_news}
\end{table}

Within detected IBs associated with scheduled announcements, we note the prevalence of announcements related to the US economy. To check for over- and under- representation of the subpopulations we performed an hypergeometric test. We find that USD-related announcements are heavily over represented compared to a random matching (p-values lower than $10^{-12}$), while all the other geographic origins are under represented. We remark that this test does not take into account the time-of-day distribution of the announcements. For example, most announcements concerning JPY are released  between 23:00 and 8:00 London time, while our test assumes the same probabilities irrespectively of the time of day. It appears nevertheless that, at least for these currency pairs, US-related economic announcements are far more likely to result in an IB compared to announcements on other countries/currency.

\begin{figure}[tb]
\centering
\includegraphics[width=0.9\textwidth]{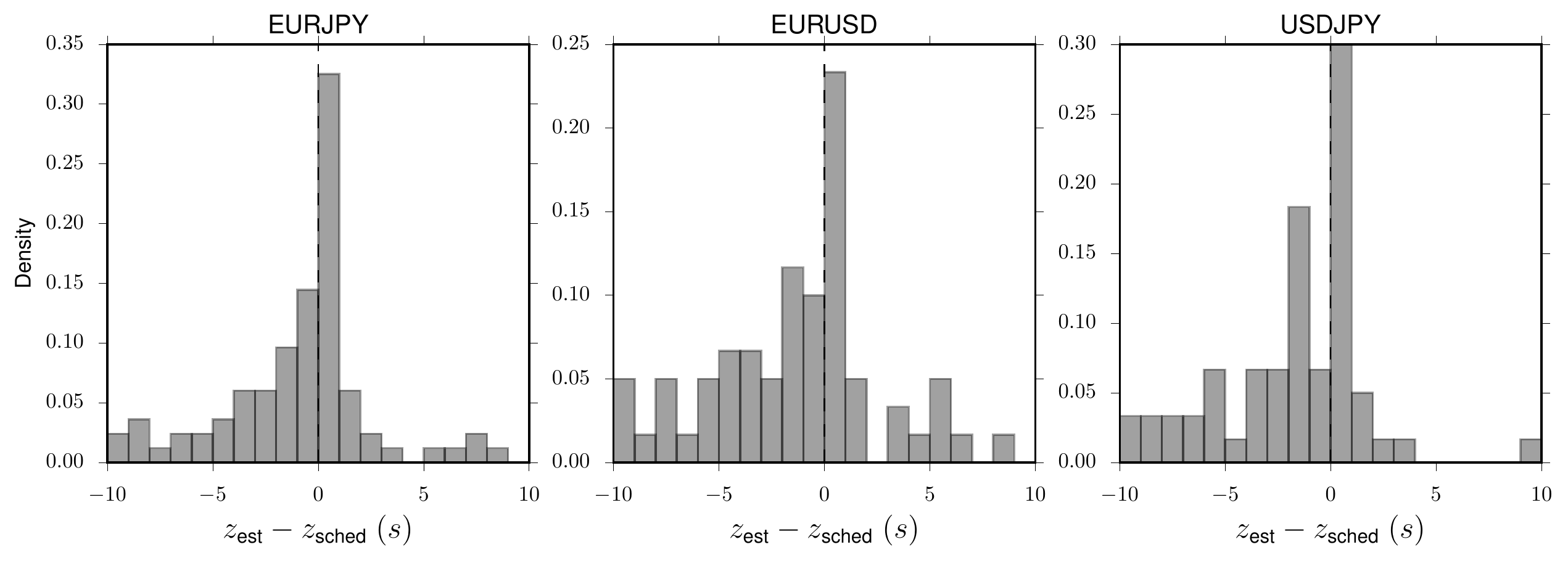}
\caption{Histogram of the time differences between the detected IB time resulting from our procedure and the scheduled release time of economic announcements for which a match was found.}
\label{fig:dz_news}
\end{figure}

It is interesting to note that the IBs do not always follow the scheduled announcement, but sometimes precede it. Figure \ref{fig:dz_news} shows the distribution of the time difference between the estimated IB starting time $z$ and the scheduled release time for the announcement for which a match was identified. The peak of the density is within a second after the scheduled time with a few cases of lags more than one second. However, there exist a substantial number of IBs that have started well before the announcement time, indicating the preparation of the market participants for the upcoming release. In the following section we will observe such anticipation in the dynamics of the spread that reflect the risk perception of the market makers.

%================================================================================ 

Bid-ask spread is an important metric of the trading costs and the liquidity of the market. Being set up usually by a market makers it often reflects their anticipation of the uncertainty in price moves. It is known that right before an important announcement liquidity often ``evaporates'' from the order book leaving a wide spread and a small amount of limit orders in the first levels, as no one wants to be affected by the possible adverse price move. Here we provide a quantitative analysis for such cases.

We consider the bid-ask spread dynamics in the windows of 2 minutes before and after the detected IBs (i.e. in window $[z-120, z+120]$). In order to aggregate data from a different periods with potentially different market conditions we normalize the spread values by the average value of the spread in a 3 minutes window ending 5 minutes before the detected IB (i.e. $[z-480, z-300]$). Taking all the normalized spread series for each FX pair, we finally compute an evenly spaced time series by averaging the values in 500$ms$ bins. 

\begin{figure}[tb]
\centering
\includegraphics[width=0.9\textwidth]{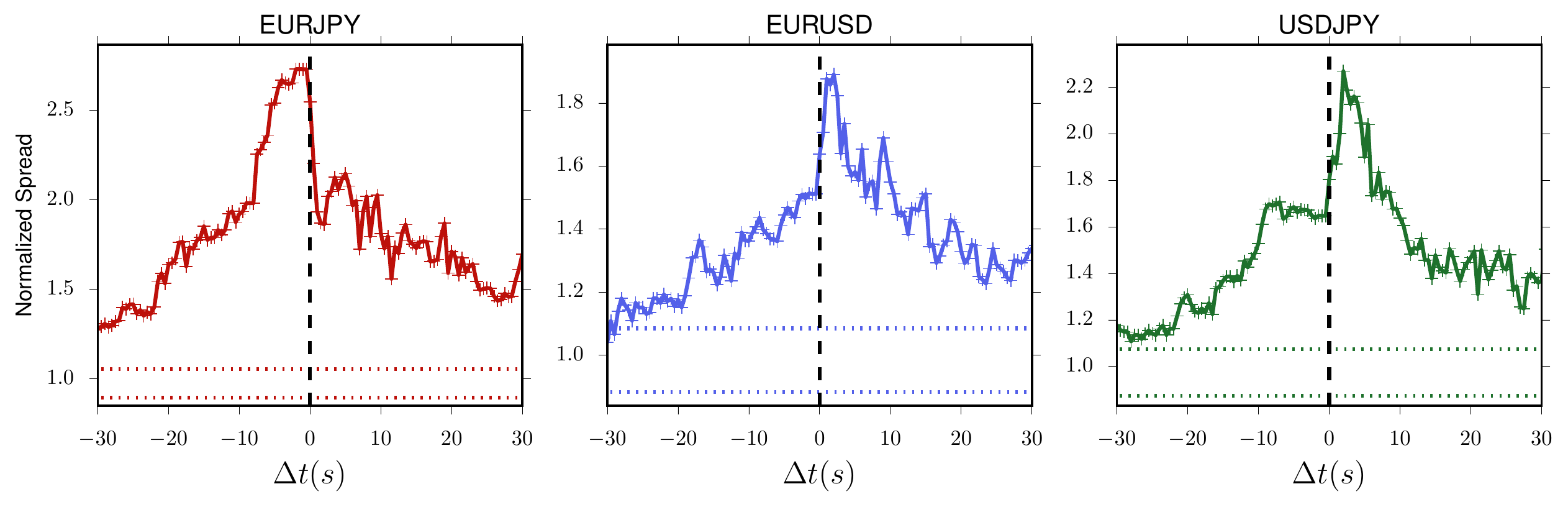}
\includegraphics[width=0.9\textwidth]{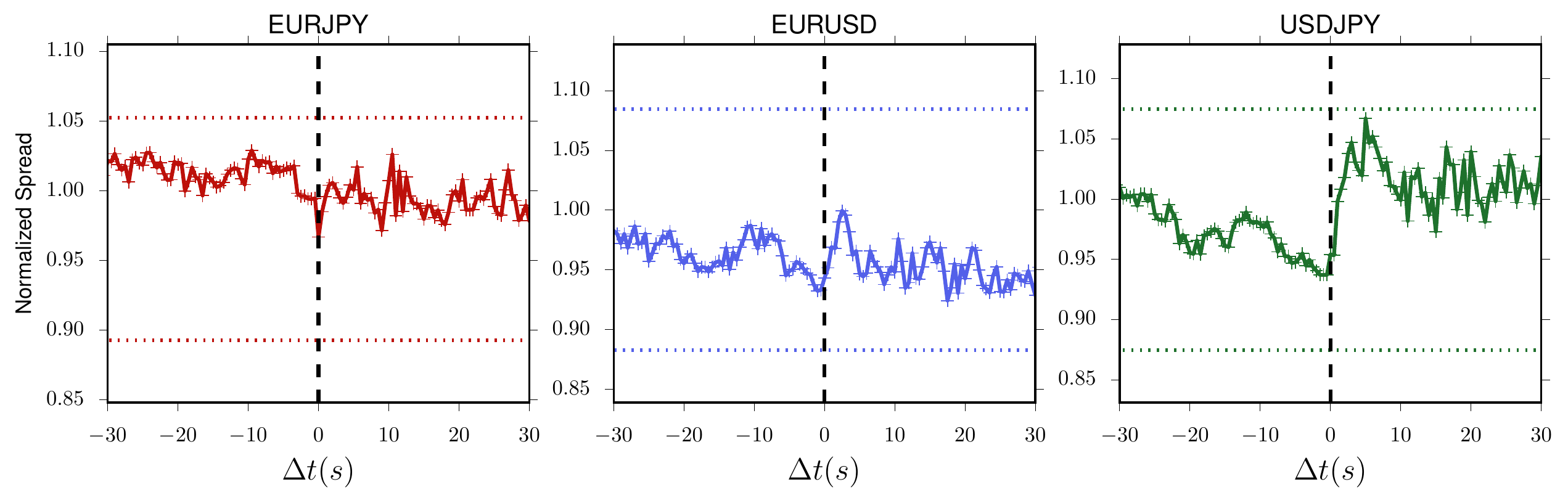}
\caption{Dynamics of the average spread around the detected IBs. Detected IBs matching (not matching) a scheduled announcement are considered in the top (bottom) panels. The dotted lines represents 5 and 95 percentiles obtained on 5000 random intervals which are at least 15 min far from an IB.}
\label{fig:spread_news}
\end{figure}

Figure~\ref{fig:spread_news} presents the average spread dynamics separately for IBs that match scheduled announcements (top panel) and IBs that do not (bottom panel). We see that in case of announcements, the spread is persistently increasing towards the detected starting time of IB (as it was discussed in previous section, often this time lies within a few seconds around the announcement times), and sharply drops right after with a following decrease. Such dynamics is rather long-lasting (of the order of tens of seconds) and appear to be statistically significantly different compared to the spread dynamics without any announcements and shocks (plotted in dashed line in Figure~\ref{fig:spread_news}).

In contrast, the spread dynamics in case of IB not associated with news does not appear to be significantly different from the average dynamics obtained on random intervals. However we can note that a few seconds right before the IB the spread tends to drop and widen right after the IB. The most pronounced this pattern is in the USDJPY currency pair. 
This observation suggests that market is not able to anticipate the arrival of these type of IBs. 

%================================================================================ 
\subsection{Intensity bursts and price jumps}

The second explanation of IBs is related to price jumps, since these could provide a possible mechanism that triggers IBs, particularly when the price jump is due to or associated with lack of liquidity. Vice versa the IBs might be a consequence of the activity of market-makers searching for a stable price level under the uncertainty of the future price moves, and a price shock or an established trend could be a resolution of such uncertainty.

\begin{figure}
\centering
\includegraphics[width=0.3\textwidth]{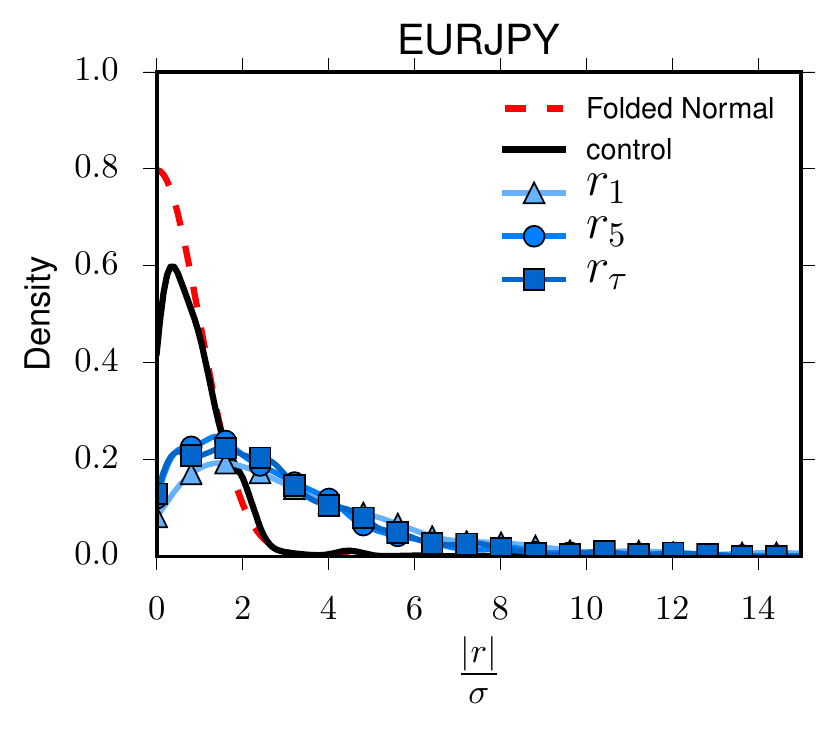}
\includegraphics[width=0.3\textwidth]{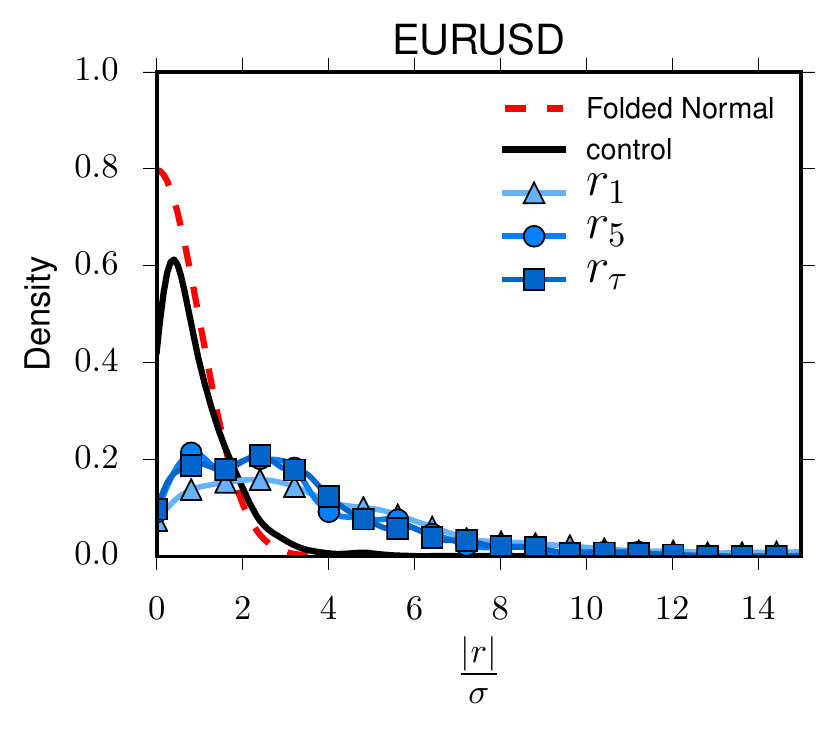}
\includegraphics[width=0.3\textwidth]{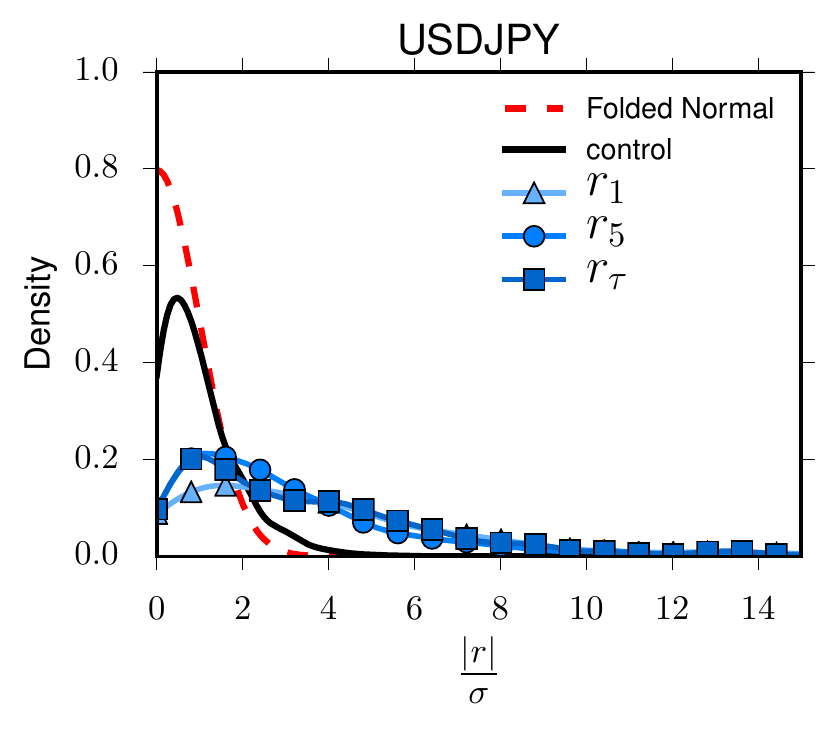}
\caption{Kernel density estimation of the distribution of the absolute normalized returns $\frac{|r_i|}{\sigmaloc}$ during an IB. For comparison also the folded Normal distribution and the empirical distribution from a control group of 1,000 randomly chosen 1 minute returns are shown. }
\label{fig:ret_dist}
\end{figure}

To motivate the analysis of the relation between IBs and large price movements, in Figure \ref{fig:ret_dist} we plot the density estimation of the distribution of the absolute normalized returns associated with detected IBs. We consider three time scales for returns, one minute, five minutes, and $\tau$ minutes (see below for details). The corresponding empirical means and standard deviations as well as excess kurtosis are reported in Table~\ref{tab:stat_dist}. We also compare it with a control distribution obtained from the random sample. It is clear that the conditional distribution has much heavier tail with a large probabilities of extreme returns. The main conclusion is that IBs are often statistically associated with large price movements, thus it is important to consider the relation between IBs and jumps. 

\begin{table}
\centering
\input{./Tables/summary_stats_return_distrib.tex}
\caption{Mean, standard deviation, and excess kurtosis of the empirical distributions of the absolute normalized returns $\frac{|r_i|}{\sigmaloc}$ during an IB. Theoretical values of the standard folded normal distribution are also reported for comparison.}
\label{tab:stat_dist}
\end{table}

There is a vast econometric literature on price jumps and their detection (\cite{andersen2007no, Lee:2008ip, bollerslev2009discrete, ait2009testing} to name only a few), we refer to~\citet{bormetti_cojumps} for literature survey and discussion about modeling co-jumps within the Hawkes process framework. 
Here, we are mostly interested in intraday abnormal returns on the same lines as \cite{Lee:2008ip} and \cite{joulin2008stock}. More specifically we consider the series of absolute midprice log-returns $|r_i| = |\log(P_i/P_{i-1})|$ on a given time scale and we say that the $i$th return $r_i$ is a $\theta\sigma$-jump if the volatility-normalized return is larger than a certain fixed threshold $\theta$:
\begin{equation}
\label{eq:pjump}
\frac{|r_i|}{\sigmaloc}> \theta 
\end{equation}
where $\sigmaloc$ is an estimate of the local volatility at time $t_i$. Here we have employed the Realized Bipower Variation~\citep{Lee:2008ip}:
\begin{equation}
\label{eq:rbv}
\sigma^2_{\mbox{\tiny RBV}} (t_i)= \frac{1}{K-2} \sum_{j=i-K+2}^{i-1} \left\vert \log \left(\frac{P(t_j)}{P(t_{j-1})}\right)\right\vert \left\vert \log \left(\frac{P(t_{j-1})}{P(t_{j-2})}\right)	\right\vert
\end{equation}
using one minute returns ($t_i-t_{i-1}=60s$) on a $K=120$ minutes window. This estimator has the advantage of being relatively robust to presence of the tail events within the estimation window. However, we found other estimators such as the Realized Variance to yield substantially similar results for our purposes.

We estimate the probability of observing a price jump given the presence of an IB at time $z$. For this we analyze price dynamics around $z$ on three time scales: 1 minute (computing return $r_1$ between $z-10s$ and $z+50s$), 5 minutes ($r_5$ is computed between $z-50s$ and $z+250s$) and a time-scale which is defined by the relaxation time~$\tau$ of the IB (we calculate return $r_\tau$ between $z-\frac{\tau}{6}$ and $z+\frac{5\tau}{6}$). For the definition of the $\theta\sigma$-jumps~\eqref{eq:pjump} we use a suitably rescaled volatility~\eqref{eq:rbv} which is originally calculated on 1 minute intervals: namely we use $\sigma_{\mbox{\tiny loc},1}=\sigmaloc$, $\sigma_{\mbox{\tiny loc},5}=\sqrt{5}\sigmaloc$ and $\sigma_{\mbox{\tiny loc},\tau}=\sqrt\tau\sigmaloc$ respectively.

Our results are reported in Table \ref{tab:jumps_aft_shck}, which presents the fraction of times where a price jump~\eqref{eq:pjump} is observed when an IB is detected by our procedure. It is seen that a large fraction of IBs is accompanied by significant price moves: for example 25\%--30\% of all IBs are accompanied with $5\sigma$-jumps on 1 minute time scale, and often these price moves do not immediately mean-revert: 12\%--20\% of 5-minute returns exceed $5\sigma$ threshold.

\begin{table}
\centering
\input{./Tables/PjumpsGivenShock.tex}
\caption{Fraction of times a price jump as defined in \eqref{eq:pjump} is found in correspondence of an IB. Results for different return horizons and three values of the threshold $\theta$ are reported.}
\label{tab:jumps_aft_shck}
\end{table}

We have also explored what indications can the model parameters give on the likelihood that an IB is accompanied by a price jump. For this we have performed a logistic-regression analysis, presented in~\ref{app:jumps}. We found that the probability of obtaining a price jump simultaneously with an IB is controlled mostly by the amplitude $\alpha$ of the IB. Surprisingly, the relaxation time $\tau$ does not seem to be a significant contributor, while adding the branching ratio $n$ as regressor improves the performance of the classifier. Overall, the classifier using both $\alpha$ and $n$ is quite reliable if it predicts the presence of a jump, while it suffers of a high false negative rate. 

So far we have investigated the presence of a price jump given a nearby IB. Now we consider the reverse question, namely whether the price jumps are accompanied by an intensity shocks. We again consider absolute one minute returns normalized by the local volatility estimated using Eq.~\eqref{eq:rbv}. 
Table \ref{tab:jump_vs_shocks} shows the fraction of these jumps that match an IB within $\pm 5$ min. Depending on threshold level $\theta$, from 20 to 40\% of all price jumps correspond to an IB as detected by our procedure. 

\begin{figure}
\centering
\includegraphics[width=0.3\textwidth]{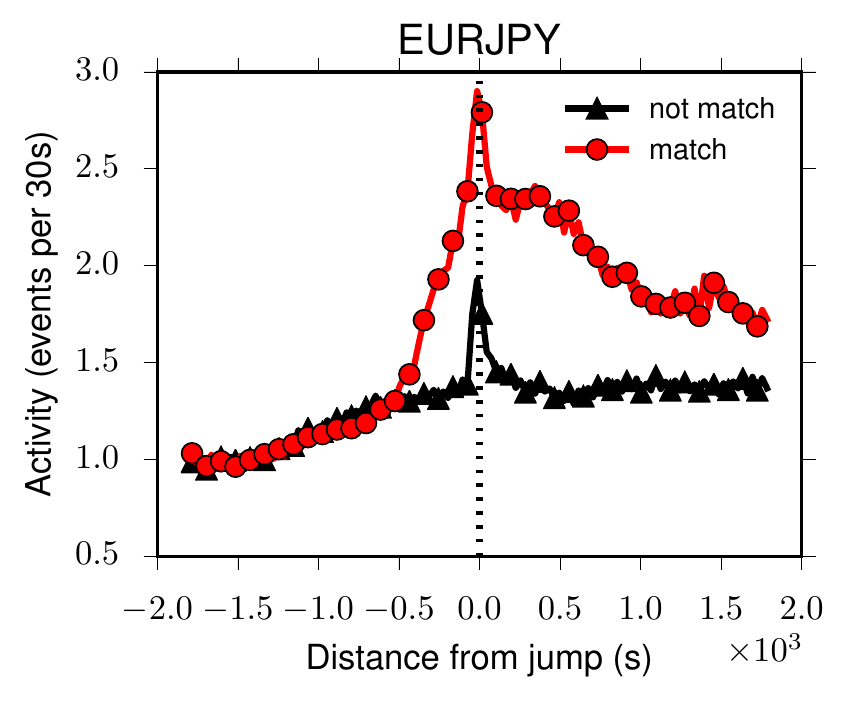}
\includegraphics[width=0.3\textwidth]{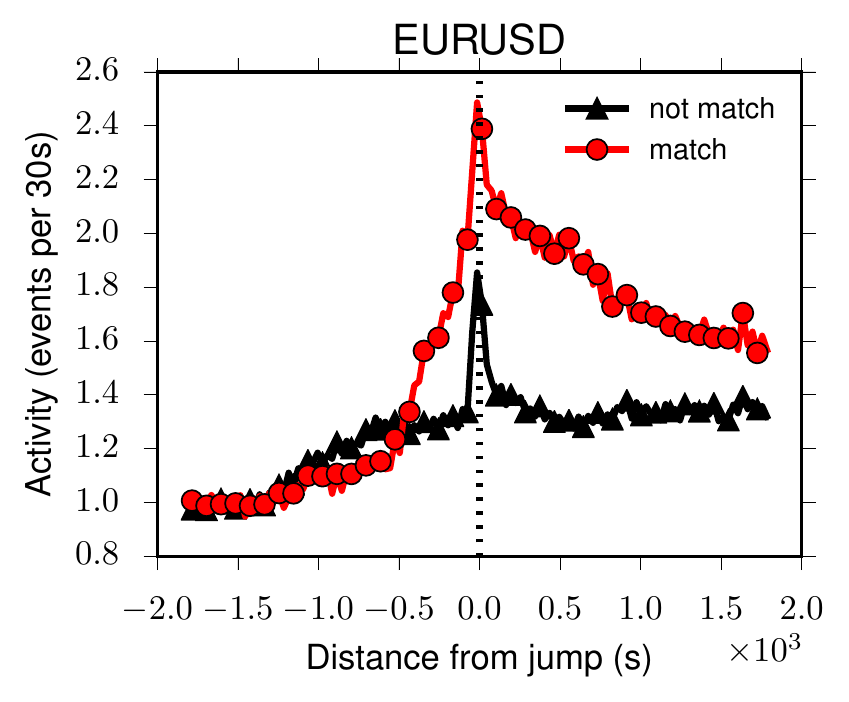}
\includegraphics[width=0.3\textwidth]{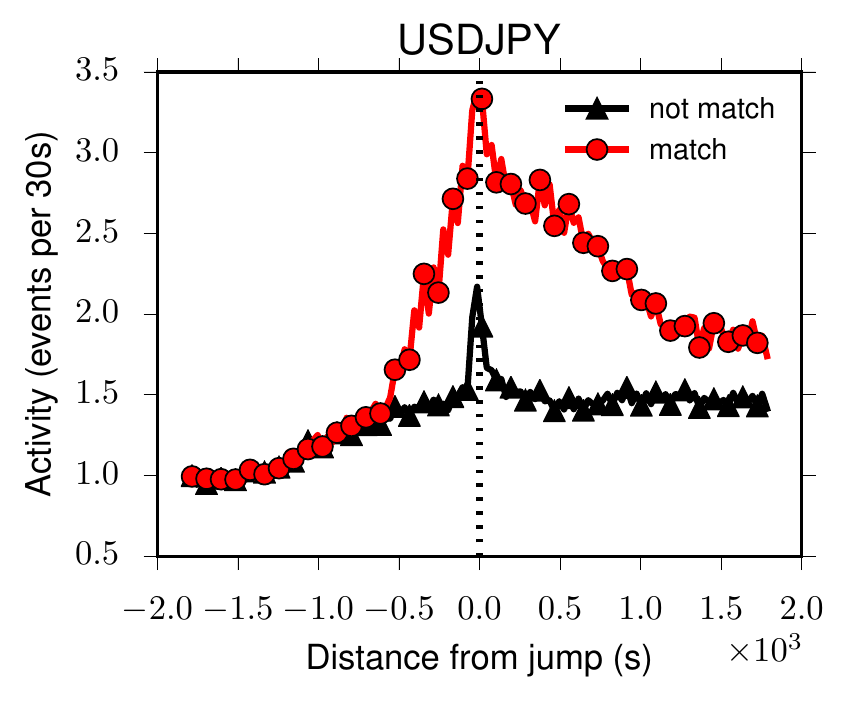}

\includegraphics[width=0.3\textwidth]{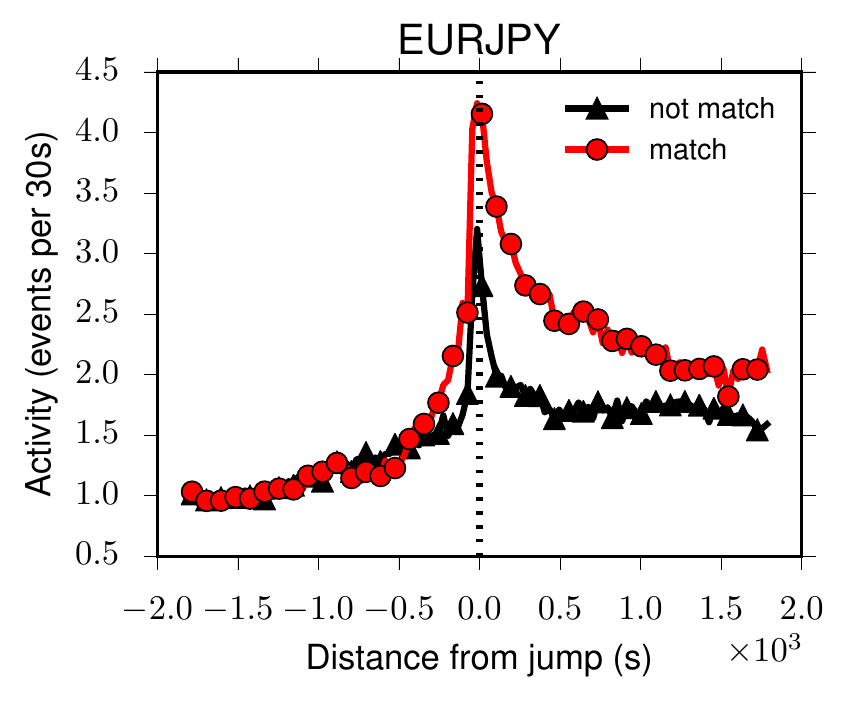}
\includegraphics[width=0.3\textwidth]{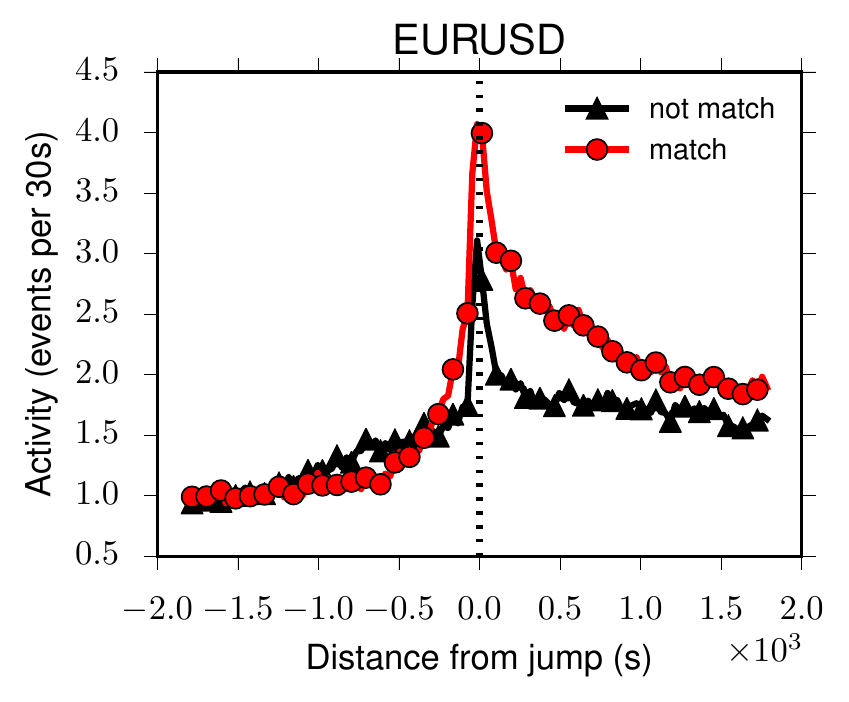}
\includegraphics[width=0.3\textwidth]{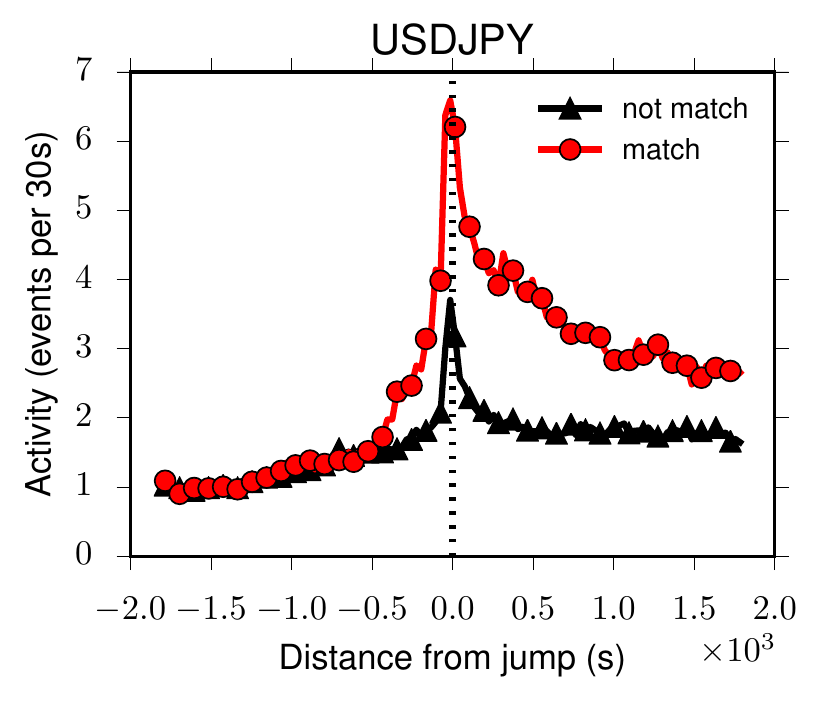}

\caption{Mean normalized activity pattern around price jumps for jumps that match (do not match) with activity IBs detected by our procedure. The window is centered on price jump locations and is 1h wide. Before averaging, the activity is normalized by the mean activity in the first 10 minutes of the window. Top panels correspond to $3<\theta <5$, bottom panels - to $\theta \geq 5$.}
\label{fig:match_vs_no}
\end{figure}

Figure \ref{fig:match_vs_no} presents the average activity around the price jumps for jumps that match an IB and for those that do not. We note that on average price jumps are associated with an increase of activity, though clearly not all of these activity spikes were identified as IB by our procedure (black lines in Figure): those activity spikes were not significantly different from the background endogenous process in terms of magnitude. Furthermore, the relaxation of activity after the price jump appears to be much faster for price jumps without IB that for those associated with an IB. It is interesting to note, that activity increases towards the price jumps on average in all cases. This suggests that most of detected price shocks are not exogenous, but either endogenously generated by the feedback mechanisms of the system, or anticipated exogenous shocks such as scheduled macro economical announcements.

\begin{table}
\centering
\input{./Tables/fraction_jumps_ret1_match_shock_tol300.tex}

\caption{Fraction of price jumps that match an IB within $\pm 5$ min. }
\label{tab:jump_vs_shocks}
\end{table}

One important result of our empirical analysis is that a significant fraction of IBs are not associated to macroeconomic news and jumps. These abrupt increases in high frequency volatility appear to be endogenously generated and not related to large price movements. Despite the fact that a full understanding of these phenomena is beyond the scope of this paper, it is interesting to present a possible explanation for them, even if based on visual inspection of selected cases.

By looking at the bid-ask dynamics around these IBs not associated to price jumps or news we observe that, while before the start of the detected IB the midprice changes are relatively rare on the considered time scale, after it there is an intense readjustment process of best quotes, without a significant net price movement. This quote flickering is frequently observed in many other IBs not associated to news or price jumps. Although we do not have a full explanation of this behavior, we note that it resembles a known strategy used by High Frequency Traders termed {\it quote stuffing} and consisting in flooding the market with huge numbers of orders and cancellations in rapid succession. 
%================================================================================ 
\section{Conclusions}
\label{sec:conc}

In this paper we presented a novel procedure to detect intensity bursts in activity of a point process within a Hawkes processes framework. The ability to separate genuine external perturbation to the system from bursts generated by internal feedback mechanisms and correlations is a key feature of our procedure. From extensive numerical tests on both synthetic and real data we find that our procedure can be very effective in identifying sudden and short lived increases in activity that are not compatible with the endogenous or "normal" dynamics.  

We applied the method to high frequency financial data describing the midprice dynamics in FX markets. We found a large number of IBs, both idiosyncratic to a specific rates and common to the three rates. Some of the IBs can be temporally related to price jumps and macroeconomic announcements. In the latter case we were able to compare the scheduled timing of the news with the inferred time of the start of the IB. In the former case we found that a large fraction of large jumps are associated with  IBs. Interestingly a significant fraction of IBs appear to be unrelated with jumps and news, opening the question of the possible market events associated to them.    

As a final remark we observe that the extension of the Hawkes model for incorporation of the explicit impact of IBs can be very relevant for mitigating the bias in the estimation of the branching ratio $n$ in systems where exogenous shocks might be present. Indeed, we showed that when such shocks are not properly accounted for, conclusions based on standard Hawkes models can severely overestimate the branching ratio.

In conclusion, we stress that, although our methodology was developed with financial applications in mind, it can be applied wherever the dynamics is naturally bursty and correlated and the influence of sudden external shocks is relevant. Examples include tweets, email, web clicks, arrival of customers and many others.

%================================================================================ 
\section*{Acknowledgements}

We are grateful to the Chair of Entrepreneurial Risks of ETH Z\"urich and specifically to Prof.\ Didier Sornette for fruitful discussions and financial support. We also thanks Spencer Wheatley and Prof. Frederic Abergel for useful discussions.

\bibliography{bib}
\bibliographystyle{chicago}

\appendix
\clearpage

%================================================================================  
%================================================================================
\section{Details of likelihood optimization}
\label{app:like}
The log-likelihood \eqref{eq:loglike} for model \eqref{eq:news_model} with endogenous kernel of the form $\phi(t) = n h(t)$ and exogenous kernels of the form $\phi_S(t) = \alpha q(t)$ and given the observations $\lbrace t_i \rbrace_{i=1,\dots,N} \in [0, T]$ reads
\begin{equation}
\label{eq:like_spec}
\begin{split}
\log \like(\mu, n, \psi, \lbrace \alpha_k, z_k, \xi_k \rbrace_{k=1,\dots,M})  = &-\mu T - n H_1(\psi) - \sum_{k=1}^M \alpha_k K_1(z_k, \xi_k) \\
&+ \sum_{t_i} \log \left( \mu + n H_2(\psi; t_i) + \sum_{k=1}^M \alpha_k K_2(z_k, \xi_k; t_i) \right)
\end{split}
\end{equation}
where $\mu$ is the baseline intensity parameter, ($n, \psi$) denotes the endogenous kernel parameters, ($\alpha_k, z_k, \xi_k$) are the parameters of the $k$-th IB, $M$ is the total number of IBs, and 
\begin{align}
H_1 &= \sum_{t_i} \left( \tilde{h}(T-t_i) -\tilde{h}(0) \right)\\
H_2 &= \sum_{t_j < t_i} h(t_i-t_j)\\
K_1(k) &= \tilde{q}(T-z_k) - \tilde{q}(0)\\
K_2(k; t_i) &= q(t_i - z_k) 
\end{align}
with $\tilde{g}$ denoting the antiderivative of the function $g$. We note that model \eqref{eq:hawkes_def} is recovered when $M=0$.
As noted in \cite{Filimonov:2015fm}, in the optimization of \eqref{eq:like_spec}, one parameter can be obtained from the relation
\begin{equation}
n \frac{\partial \log\like}{\partial n} + \mu 	\frac{\partial  \log\like}{\partial \mu} + \sum_k f_k \frac{\partial  \log\like}{\partial \alpha_k} = -\mu T -n H_1 - \sum_{k=1}^M \alpha_k K_1(z_k, \xi_k) + N
\end{equation}
which holds at the optimum. Hence, finding the maximum of \eqref{eq:like_spec} is equivalent to minimize
\begin{equation}\label{eq:cost_fun}
\begin{split}
G(n, \psi, \lbrace \alpha_k, z_k, \xi_k \rbrace_{k=1,\dots,M}) = &- \sum_{t_i} \log \left[ \frac{N}{T} + n \left(H_2(\psi;t_i) - \frac{H_1(\psi)}{T} \right) \right.\\
&\left.+ \sum_{k=1}^M \alpha_k \left(K_2(z_k, \xi_k; t_i) - \frac{K_1(z_k, \xi_k)}{T}\right) \right]
\end{split}
\end{equation}

When $M=0$ we perform the optimization using standard optimizers such as the L-BFGS-B method \citep{byrd1995limited}.  Since $G$ is usually not convex, multiple starting points are tried in order to improve the chances of finding the global optimum \citep{Filimonov:2015fm}.

When adding one IB term we use a subordination procedure to perform the optimization in a similar spirit to \cite{Filimonov:2015fm}. 
Specifically, we proceed as follows. Let us indicate with $\theta_1$ all the parameters of the model except the IB location $z_1$. Then, we separate the optimization into the two step
\begin{equation}\label{eq:step1}
\hat{\theta}_1 = \argmin_{\theta_1} S(\theta_1) 
\end{equation}
with
\begin{equation}\label{eq:step2}
S(\theta_1) = \min_{z_1 \in \lbrace t_i \in W_1\rbrace} G(\theta_1, z_1)
\end{equation}

Using the guess $\bar{z}_1$ from the pre-estimation procedure, we minimize $G$ with respect to $\theta_1$ while keeping fixed $z_1=\bar{z}_1$. Then in step \eqref{eq:step2} we update the estimate of $z_1$ while keeping $\theta_1$ fixed. We perform the optimization \eqref{eq:step1} using standard quasi-newton algorithms, while the optimization with respect to the IB location $z_1$ can be performed with a direct search over the values $t_i \in W_1$, as repeated evaluation of \eqref{eq:cost_fun} for different values of $z_1$ is very cheap computationally. 

When we add another IB to the model we proceed in a similar fashion. We optimize again all the parameters except the first IB location and we separate as before the optimization over $z_2$ from the one over the other parameters $\theta_2$.

%================================================================================

\section{Appendix to Section \ref{sec:model_validation}}\label{sec:appendix_validation}

\begin{table}[htb]
\centering
\small
\input{./Tables/one_shock_fixed_size_5000_n_0.3.tex}
\input{./Tables/one_shock_fixed_size_5000_n_0.5.tex}
\vspace{5pt}

\input{./Tables/one_shock_fixed_size_5000_n_0.7.tex}
\input{./Tables/one_shock_fixed_size_5000_n_0.9.tex}
\caption{Percentage of correctly classified IBs for different combinations of the true IB parameters $(f, \tau)$. In this experiment we do not use the pre-identification algorithm to limit the search space for the IB location but instead we directly provide the correct search interval. The results refer to a sample size of roughly 5,000 events.}
\label{tab:one_shock_fixed}
\end{table}

\begin{figure}[htb]
\small
\centering
\includegraphics[width=0.48\textwidth]{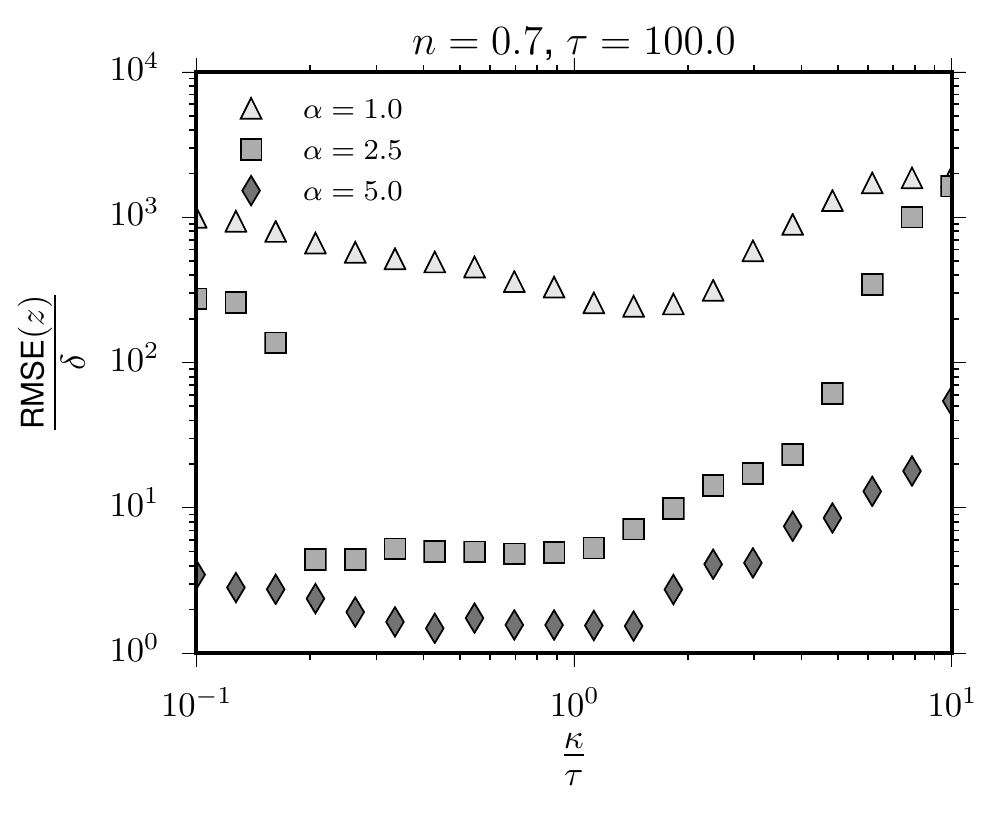}
\includegraphics[width=0.48\textwidth]{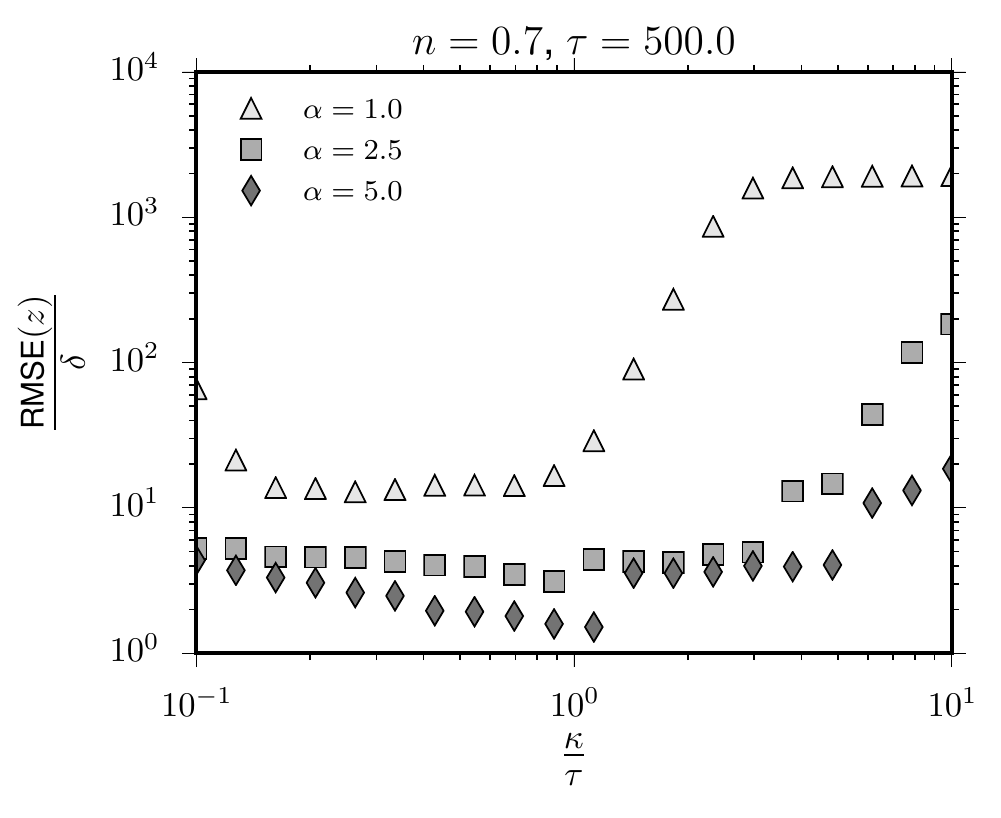}
\caption{Root Mean Squared Error on the IB location $z$ relative to the typical inter-event time  $\delta=\frac{T}{N}$ as a function of $\frac{\kappa}{\tau}$. In the above figures $T=3,600$, roughly $5,000$ events, and $n=0.7$. 100 simulations were performed for each combination $(\alpha, \tau)$.  }
\label{fig:mse_z}
\end{figure}

\begin{table}
\small
\centering
\input{./Tables/one_shock_size_5000_n_0.3_nbase.tex}
\input{./Tables/one_shock_size_5000_n_0.3_nbest.tex}
\vspace{5pt}

\input{./Tables/one_shock_size_5000_n_0.7_nbase.tex}
\input{./Tables/one_shock_size_5000_n_0.7_nbest.tex}
\caption{Average values of the branching ratio $n$ of the endogenous kernel, obtained from the base Hawkes model (left) and from the best model selected by our procedure (right) when the simulated time series has a IB with parameters $(f,\tau)$. Here we present the case where the true values are $n=0.3$ (top) and $n=0.7$ (bottom). 
}
\label{tab:n_vs_n}
\end{table}

\begin{table}[tbp]
\centering
\input{./Tables/two_shocks_appendix.tex}
\caption{Results of the tests on simulation with two IBs for the $n=0.3$, $n=0.5$, and $n=0.9$ cases. All quantities are expressed in percent. C stands for close, F for far, L for large, and S for small (see main text for details). }
\label{tab:two_shocks_app}
\end{table}

%================================================================================ 
\section{Errors on IB parameters}\label{sec:appendix_parameters}
\FloatBarrier
\begin{table}[htb]
\small
\centering
\input{./Tables/err_alpha_size_5000_n_0.3.tex}
\input{./Tables/err_alpha_size_5000_n_0.5.tex}
\vspace{5pt}

\input{./Tables/err_alpha_size_5000_n_0.7.tex}
\input{./Tables/err_alpha_size_5000_n_0.9.tex}
\caption{Relative Root Mean Squared Error (\%) on the $\alpha$ parameter obtained with our procedure on simulations where a single IB is present. One hundred simulation were generated for each combination of $n$, $f$, and $\tau$. The MSE is shown only for cases where the IB was detected in at least 50 out of 100 repetitions. These tables refers to a sample size of approximately 5000 events.}
\label{tab:mse_alpha_5000}
\end{table}

\begin{table}
\small
\centering
\input{./Tables/err_alpha_size_10000_n_0.3.tex}
\input{./Tables/err_alpha_size_10000_n_0.5.tex}
\vspace{5pt}

\input{./Tables/err_alpha_size_10000_n_0.7.tex}
\input{./Tables/err_alpha_size_10000_n_0.9.tex}
\caption{Relative Root Mean Squared Error (\%) on the $\alpha$ parameter obtained with our procedure on simulations where a single IB is present. One hundred simulation were generated for each combination of $n$, $f$, and $\tau$. The MSE is shown only for cases where the IB was detected in at least 50 out of 100 repetitions. These tables refers to a sample size of approximately 10000 events.}
\label{tab:mse_alpha_10000}
\end{table}

\begin{table}
\small
\centering
\input{./Tables/err_tau_size_5000_n_0.3.tex}
\input{./Tables/err_tau_size_5000_n_0.5.tex}
\vspace{5pt}

\input{./Tables/err_tau_size_5000_n_0.7.tex}
\input{./Tables/err_tau_size_5000_n_0.9.tex}
\caption{Relative Root Mean Squared Error (\%) on the $\tau$ parameter obtained with our procedure on simulations where a single IB is present. One hundred simulation were generated for each combination of $n$, $f$, and $\tau$. The MSE is shown only for cases where the IB was detected in at least 50 out of 100 repetitions. These tables refers to a sample size of approximately 5000 events.}
\label{tab:mse_tau_5000}
\end{table}

\begin{table}
\small
\centering
\input{./Tables/err_tau_size_10000_n_0.3.tex}
\input{./Tables/err_tau_size_10000_n_0.5.tex}
\vspace{5pt}

\input{./Tables/err_tau_size_10000_n_0.7.tex}
\input{./Tables/err_tau_size_10000_n_0.9.tex}
\caption{Relative Root Mean Squared Error (\%) on the $\tau$ parameter obtained with our procedure on simulations where a single IB is present. One hundred simulation were generated for each combination of $n$, $f$, and $\tau$. The MSE is shown only for cases where the IB was detected in at least 50 out of 100 repetitions. These tables refers to a sample size of approximately 10000 events.}
\label{tab:mse_tau_10000}

\end{table}

%================================================================================ 
\section{Kernel misspecification}
\label{app:km}

In this section we present the result obtained by our procedure when the endogenous kernel $\phi(t)$ used to generate the data differs from the one used in the estimation.

In particular we run simulations with a single exponential kernel
\begin{equation}
\label{eq:se}
\phi_{\mbox{\tiny{SE}}} = n b e^{-b t}
\end{equation}
and a double exponential kernel
\begin{equation}
\label{eq:de}
\phi_{\mbox{\tiny{DE}}} = n \left( ab_A e^{-b_A t} + (1-a)b_B e^{-b_B t} \right).
\end{equation}

The test procedure is exactly the same as outlined in Sections \ref{sec:ss_0} and \ref{sec:ss_1}. The result presented are for a target sample size of 10,000. In the following, SE1 (SE2) will denote simulations with the single exponential as kernel and parameter $b=0.1$ ($b=1$), while DE will denote simulations with the double exponential kernel and parameters $a=0.7, b_A=2$, and $b_B=0.1$. 

In Table \ref{tab:km_0} we present the rate of false positives for simulations where no IB is present, while in Tables \ref{tab:se1}, \ref{tab:se2}, \ref{tab:de} the rate of correctly classified IBs for the SE1, SE2 and DE cases respectively. Our procedure of the IB identification appears to be robust with respect to miss-specifications of the model.

\begin{table}
\centering
\input{./Tables/ker_mis_no_shock.tex}
\caption{Percentage of false positive using Bayesian Information Criterion when the simulated model has no IBs. 1000 simulations where performed in each case. All values expressed in percent (\%).}
\label{tab:km_0}
\end{table}

\begin{table}[htb]
\small
\centering
\input{./Tables/KM_one_shock_case_SE1_n_0.3.tex}
\input{./Tables/KM_one_shock_case_SE1_n_0.5.tex}
\vspace{5pt}

\input{./Tables/KM_one_shock_case_SE1_n_0.7.tex}
\input{./Tables/KM_one_shock_case_SE1_n_0.9.tex}
\caption{Percentage of correctly classified IBs for different combinations of the true IB parameters $(\alpha, \tau)$ expressed in terms of $f=\alpha \tau$ and $\tau$. The results refer to the case SE1 and 100 simulations for each case where performed.}
\label{tab:se1}
\end{table}

\begin{table}[htb]
\small
\centering
\input{./Tables/KM_one_shock_case_SE2_n_0.3.tex}
\input{./Tables/KM_one_shock_case_SE2_n_0.5.tex}
\vspace{5pt}

\input{./Tables/KM_one_shock_case_SE2_n_0.7.tex}
\input{./Tables/KM_one_shock_case_SE2_n_0.9.tex}
\caption{Percentage of correctly classified IBs for different combinations of the true IB parameters $(\alpha, \tau)$ expressed in terms of $f=\alpha \tau$ and $\tau$. The results refer to the case SE2 and 100 simulations for each case where performed.}
\label{tab:se2}
\end{table}

\begin{table}[htb]
\small
\centering
\input{./Tables/KM_one_shock_case_DE1_n_0.3.tex}
\input{./Tables/KM_one_shock_case_DE1_n_0.5.tex}
\vspace{5pt}

\input{./Tables/KM_one_shock_case_DE1_n_0.7.tex}
\input{./Tables/KM_one_shock_case_DE1_n_0.9.tex}
\caption{Percentage of correctly classified IBs for different combinations of the true IB parameters $(\alpha, \tau)$ expressed in terms of $f=\alpha \tau$ and $\tau$. The results refer to the case DE and 100 simulations for each case where performed.}
\label{tab:de}
\end{table}

%================================================================================ 
\section{Price jumps classification}
\label{app:jumps}

We now investigate whether the IB parameters $\alpha$ and $\tau$ are good predictors for the presence of a price jump. To this end we fix $\theta=4$ and we consider a logistic model where the dependent variable is the classification jump/no jump (mapped to $\{1,0\}$ ) and the regressors are the estimates of $\alpha$ and $\tau$ or $\alpha$ and the branching ratio $n$.  For each currency pair, we fit the logistic model on a randomly chosen subsample of about 70\% of the detected IBs (later we will perform an out of sample analysis with the remaining 30\%). In Table \ref{tab:a_tau} we report the obtained coefficients as well as their standard errors for EURUSD. The results for the other pairs are very similar. We note immediately that the $\tau$ parameter does not appear to bring any contribution to the classification and indeed Akaike criterion selects the model with only $\alpha$ as regressor. When the variable $n$ is used in place of $\tau$ as a regressor we obtain a significant coefficient for $n$, indeed we find that this model is favored by AIC over the oneusing only $\alpha$ for all three pairs.

\begin{table}
\centering
\input{./Tables/logistic_alpha_tau_EURUSD.tex}
\caption{Obtained parameters value and associated errors when $\alpha$ and $\tau$ (top) or $\alpha$ and $n$ (bottom) are used as regressors. Results shown are for the EURUSD pair.}
\label{tab:a_tau}
\end{table}

\begin{figure}
\centering
\includegraphics[width=0.3\textwidth]{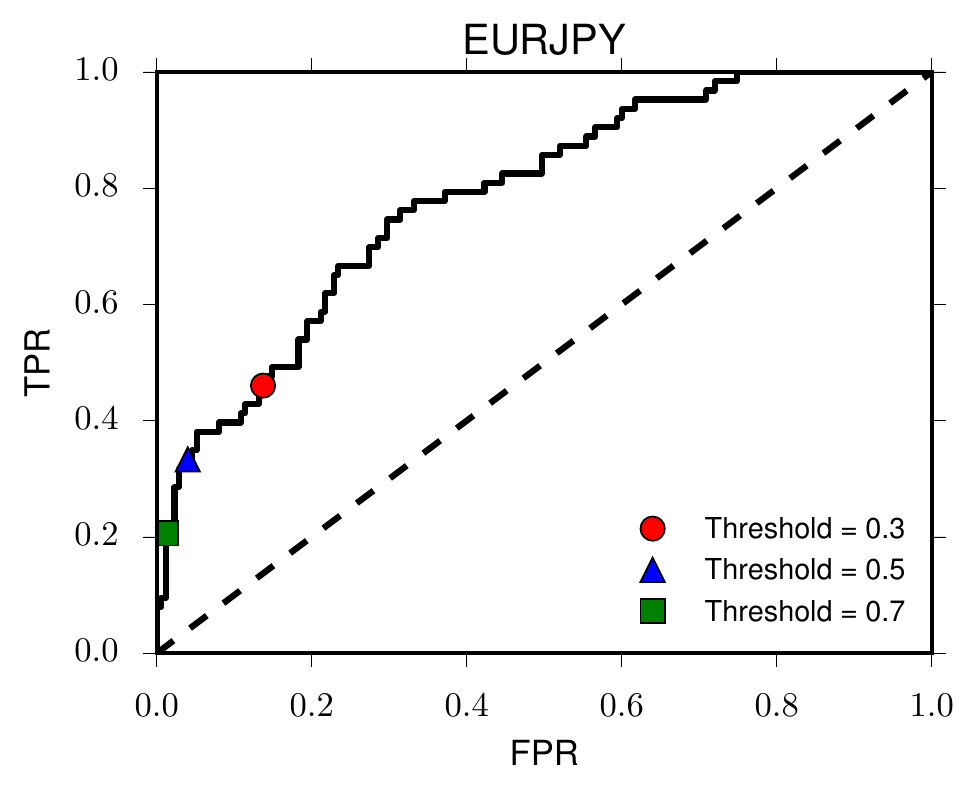}
\includegraphics[width=0.3\textwidth]{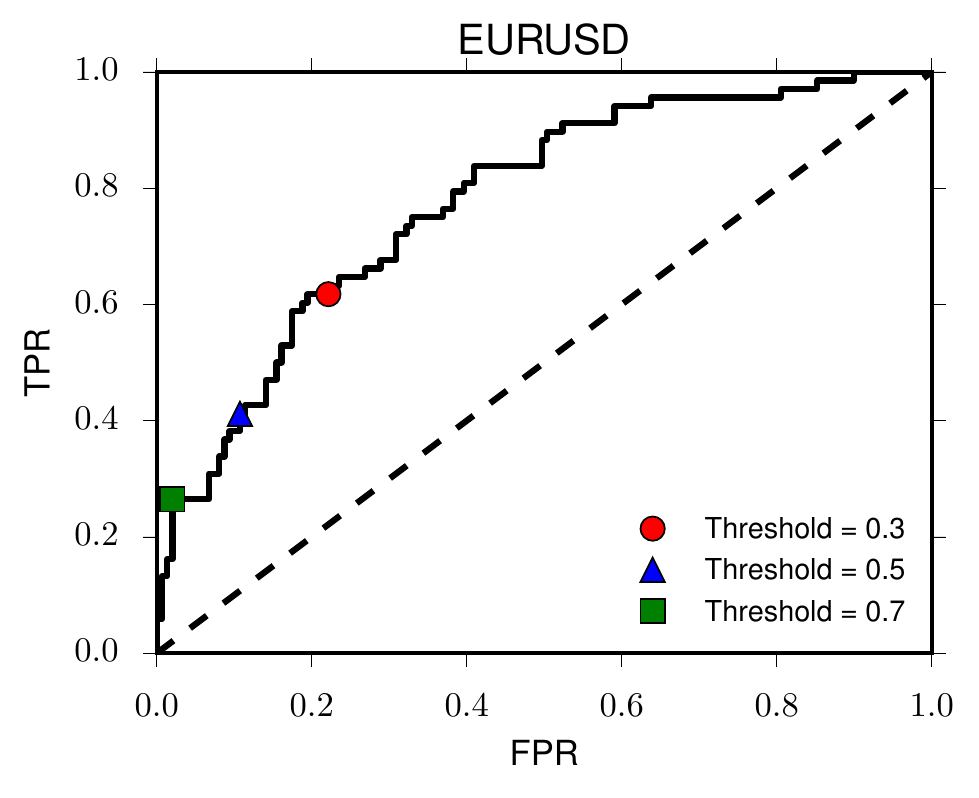}
\includegraphics[width=0.3\textwidth]{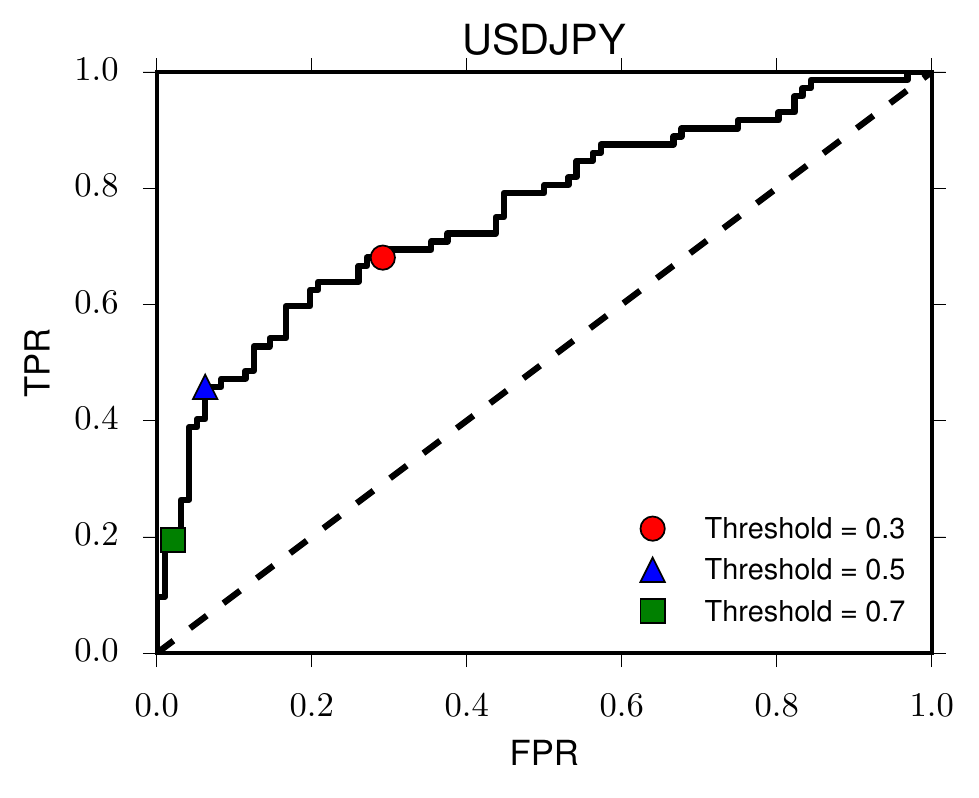}
\caption{ROC curves for the logit classifier that uses $\alpha$ and $n$ to predict whether a price jump ($\theta=4$) will be present together with the IB.}
\label{fig:roc}
\end{figure}

In Table \ref{tab:indicators} we report some performance metrics obtained by using the fitted models to predict the presence of jump in price on the remaining 30\% of the sample, using $0.5$ as threshold to predict a price jump. Overall, the prediction yielded has fairly high accuracy and precision but suffer from a high false negative rates. For completeness, in Figure \ref{fig:roc} we plot the ROC curves obtained on the test samples. Overall, the classifier based on the parameters $\alpha$ and $n$ of IB is quite reliable if it predicts the presence of a jumps, while it can lead to significant errors if it predicts that the jump is not there. 

\begin{table}
\centering
\input{./Tables/indicators.tex}

\caption{Out of sample performance metrics for the logistic model with $\alpha$ and $n$ as regressors. True Positive Rate (TPR), True Negative Rate (TNR), Precision (True Positives/Predicted Positives) and Accuracy are reported.}
\label{tab:indicators}
\end{table}

\end{document}

%% file: Tables/FPR_noshocks.tex
\begin{tabular}{lrrrrrrrr}
\toprule
& \multicolumn{4}{c}{$n$}   \\
size & 0.3 & 0.5 & 0.7 & 0.9  \\
\midrule
1000  & 0.7 & 0.5 & 1.3 & 1.0  \\
2000  & 0.4 & 0.4 & 0.7 & 0.6  \\
5000  & 0.0 & 0.3 & 0.2 & 0.2  \\
10000 & 0.1 & 0.0 & 0.3 & 0.2  \\
\bottomrule
\end{tabular}

%% file: Tables/summary_comb_min2000.tex
\begin{tabular}{lcccc}
\toprule
{} &  \# shocks &  Shocks per window &  Fraction no shock &  Max shocks \\
\midrule
EURJPY &        792 &               0.51 &               0.61 &           5 \\
EURUSD &        720 &               0.46 &               0.63 &           4 \\
USDJPY &        558 &               0.36 &               0.70 &           3 \\
\bottomrule
\end{tabular}

%% file: Tables/common2.tex
\begin{tabular}{lccc}
\toprule
{} &  \# common &  \% pair 1 &  \% pair 2 \\
\midrule
EURJPY/USDJPY &        252 &      31.82 &      45.16 \\
EURUSD/EURJPY &        324 &      45.00 &      40.91 \\
EURUSD/USDJPY &        161 &      22.36 &      28.85 \\
\bottomrule
\end{tabular}

%% file: Tables/ttest_common_noncommon_min2000.tex
\begin{tabular}{lrrrr}
\toprule
{} &  t-statistic &  $p$-value &  KS-statistic &  $p$-value \\
\midrule
2 vs 1 &         0.66 &       0.51 &         0.094 &     0.0011 \\
3 vs 1 &          8.1 &      2e-15 &           0.3 &    1.3e-21 \\
3 vs 2 &          7.3 &    8.1e-13 &          0.27 &    2.1e-16 \\
\bottomrule
\end{tabular}

%% file: Tables/summary_news_min2000.tex
\begin{tabular}{lrrr}
\toprule
{} &  \# matches &  \% news detected &  \% IBs related to news \\
\midrule
EURUSD &         118 &             10.35 &                      16.39 \\
EURJPY &         137 &             12.02 &                      17.30 \\
USDJPY &         121 &             10.61 &                      21.68 \\
\bottomrule
\end{tabular}

%% file: Tables/fraction_jumps_ret1_match_shock_tol300.tex
\begin{tabular}{lrrr}
\toprule
{} &  EURJPY &  EURUSD &  USDJPY \\
\midrule
$\theta = 3$ &    0.23 &    0.22 &    0.18 \\
$\theta = 4$ &    0.32 &    0.33 &    0.26 \\
$\theta = 5$ &    0.42 &    0.44 &    0.34 \\
\bottomrule
\end{tabular}

%% file: Tables/ker_mis_no_shock.tex
\begin{tabular}{lrrrrrrrrrrrr}
\toprule
{}& \multicolumn{3}{c}{DE} & \multicolumn{3}{c}{SE1} &\multicolumn{3}{c}{SE2} \\
$n$ &  0.3 &  0.5 &  0.7 &  0.9 &  0.3 &  0.5 &  0.7 &  0.9 &  0.3 &  0.5 &  0.7 &  0.9 \\
\midrule
FPR &  0.0 &  0.0 &  0.5 &  0.8 &  0.1 &  0.0 &  0.0 &  0.5 &  0.0 &  0.1 &  0.2 &  0.2 \\
\bottomrule
\end{tabular}

%% file: Tables/indicators.tex
\begin{tabular}{lrrr}
\toprule
{} &  EURJPY &  EURUSD &  USDJPY \\
\midrule
TPR       &    0.33 &    0.40 &    0.46 \\
TNR       &    0.96 &    0.89 &    0.94 \\
Precision &    0.75 &    0.63 &    0.85 \\
Accuracy  &    0.79 &    0.74 &    0.73 \\
\bottomrule
\end{tabular}